\documentclass[twoside,leqno]{article}

\pdfoutput=1

\usepackage{RR}
\usepackage[a4paper]{geometry}
\usepackage{a4wide}
\usepackage[utf8]{inputenc}
\usepackage[OT1]{fontenc} 

\usepackage[english]{babel}

\usepackage[numbers]{natbib}

\usepackage[colorlinks=true, citecolor=blue, linkcolor=blue]{hyperref}

\usepackage{amssymb, amsmath, amsthm, mathtools}
\usepackage{amsfonts, bbold, dsfont}

\usepackage{listings, lstcoq}
\usepackage[framemethod=tikz]{mdframed}

\usepackage{makecell}

\usepackage{longtable}

\theoremstyle{plain}
\newmdtheoremenv%
  [linecolor=black, roundcorner=5pt, linewidth=1pt]%
  {TextbookTheorem}{Textbook Theorem}

\newcommand{\LMF}{%
  Université Paris-Saclay, CNRS, ENS Paris-Saclay, Inria, Laboratoire Méthodes
  Formelles, 91190, Gif-sur-Yvette, France.}
\newcommand{\SERENA}{%
  a. Inria, 2 rue Simone Iff, 75589 Paris, France.\protect\\
  b. CERMICS, École des Ponts, 77455 Marne-la-Vallée, France.}
\newcommand{\LMAC}{%
  LMAC (Laboratory of Applied Mathematics of Compiègne), CS 60319, Université
  de technologie de Compiègne, 60203 Compiègne Cedex, France.}
\newcommand{\LIPN}{%
  LIPN, CNRS UMR 7030, Université Paris 13, 93430 Villetaneuse, France.}

\newcommand{\Title}{%
  A Coq Formalization of Lebesgue Integration of Nonnegative Functions}
\newcommand{\Titre}{%
  Une formalisation en Coq de l'intégrale de Lebesgue des fonctions positives}

\newcommand{\Abstract}{%
  Integration, just as much as differentiation, is a fundamental calculus tool
  that is widely used in many scientific domains.  Formalizing the mathematical
  concept of integration and the associated results in a formal proof assistant
  helps in providing the highest confidence on the correctness of numerical
  programs involving the use of integration, directly or indirectly.  By its
  capability to extend the (Riemann) integral to a wide class of irregular
  functions, and to functions defined on more general spaces than the real
  line, the Lebesgue integral is perfectly suited for use in
  mathematical fields such as probability theory, numerical mathematics, and
  real analysis.  In this article, we present the {\Coq} formalization of
  $\sigma$-algebras, measures, simple functions, and integration of
  {\nonnegative} measurable functions, up to the full formal proofs of the
  {\BeppoLeviMonotConvTh} and {\FatouLem}.  More than a plain formalization of
  the known literature, we present several design choices made to balance the
  harmony between mathematical readability and usability of {\Coq}
  theorems.  These results are a first milestone toward the formalization of
  $L^p$~spaces such as Banach spaces.
}

\newcommand{\Resume}{%
  Le calcul intégral, tout comme le calcul différentiel, est un outil
  fondamental utilisé largement dans de nombreux domaines scientifiques.  La
  formalisation de la notion mathématique d'intégrale et de ses propriétés dans
  un assistant de preuve aide à donner la plus grande confiance sur la
  correction de programmes numériques utilisant l'intégration, directement ou
  indirectement.  De part sa capacité à étendre l'intégrale (de Riemann) à une
  large classe de fonctions irrégulières, et à des fonctions définies sur des
  espaces plus généraux que la droite réelle, l'intégrale de Lebesgue est
  considérée comme parfaitement adaptée aux domaines mathématiques tels que la
  théorie des probabilités, les mathématiques numériques et l'analyse réelle.
  Dans cet article, nous présentons la formalisation en {\Coq} des tribus (ou
  $\sigma$-algèbres), des mesures, des fonctions étagées et de l'intégrale des
  fonctions mesurables positives, jusqu'aux preuves formelles complètes du
  théorème de convergence monotone de {\BL} et du lemme de Fatou.  Plus qu'une
  simple formalisation de la littérature connue, nous présentons plusieurs
  choix de design menés pour équilibrer l'harmonie entre la lisibilité
  mathématique et l'ergonomie des théorèmes {\Coq}.  Ces résultats sont un
  premier jalon vers la formalisation des espaces~$L^p$ comme espaces de
  Banach.
}

\newcommand{\Keywords}{%
  formal proof,
  Coq,
  measure theory,
  Lebesgue integration}
\newcommand{\Motscles}{%
  preuve formelle,
  Coq,
  théorie de la mesure,
  intégrale de Lebesgue}

\newcommand{\myskip}{\bigskip} 

\newcommand{\aka}{a.k.a.}

\newcommand{\eg}{e.g.}
\newcommand{\ie}{i.e.}

\newcommand{\viceversa}{\emph{vice versa}}

\newcommand{\nonconnected}{non\-connected}

\newcommand{\nondecreasing}{non\-decreasing}
\newcommand{\nonmeasurable}{non\-measurable}
\newcommand{\nonnegativ}{non\-negativ}
\newcommand{\nonnegative}{{\nonnegativ}e}
\newcommand{\nonnegativity}{{\nonnegativ}ity}
\newcommand{\nonnegativeness}{{\nonnegative}ness}
\newcommand{\nonstrict}{non\-strict}

\newcommand{\nonzero}{non\-zero}
\newcommand{\quasiliteral}{quasi\-literal}
\newcommand{\subadditivity}{sub\-additivity}

\newcommand{\monotony}{monotony} 

\newcommand{\code}[1]{\texttt{#1}}

\newcommand{\soft}[1]{\textsf{#1}}
\newcommand{\ACLtwo}{\soft{ACL2}}
\newcommand{\Coq}{\soft{Coq}}
  \newcommand{\Coquelicot}{\soft{Coquelicot}}
  \newcommand{\CoRN}{\soft{CoRN}}
  \newcommand{\Flocq}{\soft{Flocq}}
  \newcommand{\List}{\soft{List}}
  \newcommand{\Lists}{\soft{Lists}}
  \newcommand{\Logic}{\soft{Logic}}
  \newcommand{\MathClasses}{\soft{Math Classes}}
  \newcommand{\MathComp}{\soft{math-comp}}
  \newcommand{\MathCompAnalysis}{\MathComp\soft{/analysis}}
  \newcommand{\Reals}{\soft{Reals}}
  \newcommand{\Sorting}{\soft{Sorting}}

\newcommand{\HOL}{\soft{HOL}}
\newcommand{\HOLLight}{\soft{HOL Light}}
\newcommand{\HOLfour}{\soft{HOL4}}
\newcommand{\IsabelleHOL}{\soft{Isabelle/HOL}}
\newcommand{\Lean}{\soft{Lean}}
\newcommand{\Mizar}{\soft{Mizar}}
\newcommand{\PVS}{\soft{PVS}}

\renewcommand{\leq}{\leqslant}
\renewcommand{\subset}{\subseteq}
\renewcommand{\emptyset}{\varnothing}

\newcommand{\fhi}{\varphi}

\newcommand{\N}{\mathbb{N}}
\newcommand{\Z}{\mathbb{Z}}
\newcommand{\Q}{\mathbb{Q}}
\newcommand{\R}{\mathbb{R}}
\newcommand{\Rplus}{\R_+}
\newcommand{\Rbar}{\overline{\R}}
\newcommand{\Rbarplus}{\Rbar_+}
\newcommand{\C}{\mathbb{C}}
\newcommand{\F}{\mathbb{F}}
\newcommand{\calSF}{\mathcal{SF}}
\newcommand{\calSFplus}{\calSF_+}
\newcommand{\calM}{\mathcal{M}}
\newcommand{\calMplus}{\calM_+}

\newcommand{\calS}{\mathcal{S}}

\newcommand{\charac}[1]{{\mathds{1}}_{#1}}
\newcommand{\restr}[2]{{#1}_{|_{#2}}}
\newcommand{\Equiv}{\Leftrightarrow}
\newcommand{\EQUIV}{\quad\Longleftrightarrow\quad}
\newcommand{\Implies}{\Rightarrow}
\newcommand{\IMPLIES}{\quad\Longrightarrow\quad}
\newcommand{\eqdef}{\stackrel{\mathrm{def.}}{=}}
\newcommand{\EQDEF}{\,\eqdef\,}
\newcommand{\EQ}{\,=\,}
\newcommand{\LEQ}{\,\leq\,}

\newcommand{\PLUS}{\,+\,}
\newcommand{\almev}{\mathrm{a.e.}}
\newcommand{\muae}[1]{#1\,\almev}
\newcommand{\opae}[2]{\stackrel{\muae{#2}}{#1}}
\newcommand{\eqae}[1]{\opae{=}{#1}}

\newcommand{\intSFplus}{\int_{\calSFplus}}
\newcommand{\intMplus}{\int_{\calMplus}}
\newcommand{\floor}[1]{\left\lfloor #1 \right\rfloor}
\newcommand{\measurable}{{\mathrm{measurable}}}
\newcommand{\preimage}[2]{#1^{-1}(\{#2\})}
\renewcommand{\succ}{\mathrm{succ}}

\newcommand{\thm}[1]{#1 theorem} 
\newcommand{\thms}[1]{\thm{#1}s}
\newcommand{\BL}{Beppo Levi}
\newcommand{\BLt}{\thm{\BL}}
\newcommand{\BeppoLeviTh}{\BLt}
\newcommand{\BLmct}{\thm{{\BL} (monotone convergence)}}
\newcommand{\BeppoLeviMonotConvTh}{\BLmct}
\newcommand{\Cara}{Cara\-th\'eodory}
\newcommand{\Ct}{\thm{{\Cara}'s extension}}
\newcommand{\CaratheodoryTh}{\Ct}
\newcommand{\Dpl}{Dynkin $\pi$--$\lambda$}
\newcommand{\Dplt}{\thm{\Dpl}}
\newcommand{\DynkinPiLambdaTh}{\Dplt}
\newcommand{\Fl}{Fatou's lemma}
\newcommand{\FatouLem}{\Fl}
\newcommand{\FL}{Fatou--Lebesgue}
\newcommand{\FLt}{\thm{{\FL}}}
\newcommand{\FatouLebesgueTh}{\FLt}
\newcommand{\Gt}{\thm{Green's}}
\newcommand{\GreenTh}{\Gt}
\newcommand{\HK}{Henstock--Kurzweil}

\newcommand{\Ldcv}{Lebesgue's dominated convergence}
\newcommand{\Ldcvt}{\thm{\Ldcv}}
\newcommand{\LebesgueDominatedConvergenceTh}{\Ldcvt}
\newcommand{\LM}{Lax--Milgram}
\newcommand{\LMt}{\thm{\LM}}
\newcommand{\LaxMilgramTh}{\LMt}
\newcommand{\mct}{\thm{monotone class}}
\newcommand{\MonotClassTh}{\mct}
\newcommand{\TF}{Tonelli--Fubini}
\newcommand{\TFts}{\thms{\TF}}
\newcommand{\TonelliFubiniThs}{\TFts}

\newcommand{\FEM}{Finite Element Method}

\newcommand{\labelpar}[3]{%
  \parbox{#1in}{%
    \begin{center}
      \coqe{#2}\\
      (#3)
    \end{center}
  }
}
\newcommand{\labelwref}[3]{\labelpar{#1}{#2}{Section~\ref{sec:#3}}}
\newcommand{\labelwrefs}[4]{%
  \labelpar{#1}{#2}{Sections~\ref{sec:#3}, \ref{sec:#4}}}

\newcommand{\RbarComplLbl}[1]{\labelwref{#1}{Rbar_compl}{limit}}
\newcommand{\subsetComplLbl}[1]{\labelwref{#1}{subset_compl}{charac}}
\newcommand{\countableSetsLbl}[1]{\labelwref{#1}{countable_sets}{charac}}
\newcommand{\topoBasesRLbl}[1]{\labelwref{#1}{topo_bases_R}{topoR}}
\newcommand{\sumRbarNonnegLbl}[1]{\labelwref{#1}{sum_Rbar_nonneg}{sum_Rbar}}
\newcommand{\measurableFunLbl}[1]{\labelwref{#1}{measurable_fun}{meas_fn}}
\newcommand{\measureLbl}[1]{\labelwref{#1}{measure}{measures}}
\newcommand{\sigmaAlgebraLbl}[1]{%
  \labelwrefs{#1}{sigma_algebra}{sigma-alg}{cartesian_measu}}
\newcommand{\sigmaAlgebraRRbarLbl}[1]{%
  \labelwref{#1}{sigma_algebra_R_Rbar}{gen_equiv}}
\newcommand{\lintPLbl}[1]{\labelwref{#1}{LInt_p}{lint_p}}
\newcommand{\simpleFunLbl}[1]{\labelwref{#1}{simple_fun}{simple_fun}}

\usepackage{color}

\definecolor{tanIII}{RGB}{205,135,63}
\definecolor{yellowII}{RGB}{238,238,0}
\definecolor{darkOliveGreenIII}{RGB}{162,205,90}
\definecolor{turquoiseII}{RGB}{0,229,238}

\newcommand{\complColor}[1]{\colorbox{tanIII}{#1}}
\newcommand{\prelimColor}[1]{\colorbox{yellowII}{#1}}
\newcommand{\measColor}[1]{\colorbox{darkOliveGreenIII}{#1}}
\newcommand{\lintColor}[1]{\colorbox{turquoiseII}{#1}}

\newcommand{\ComplColorName}{\complColor{Brown}}
\newcommand{\complColorName}{\complColor{brown}}
\newcommand{\PrelimColorName}{\prelimColor{Yellow}}
\newcommand{\prelimColorName}{\prelimColor{yellow}}
\newcommand{\MeasColorName}{\measColor{Green}}
\newcommand{\measColorName}{\measColor{green}}
\newcommand{\LintColorName}{\lintColor{Blue}}
\newcommand{\lintColorName}{\lintColor{blue}}

\newcommand{\mydisplay}[1]{#1}
\newcommand{\myequal}{$ equals $}
\newcommand{\myeqdef}{$ is defined by $}
\newcommand{\mydisplaymaths}[1]{,\[#1.\]}
\newcommand{\mybreakmaths}{}
\newcommand{\mybreak}{\clearpage}

\RRdate{April 2021}

\RRversion{2}
\RRdater{November 2021}

\RRauthor{%
  Sylvie Boldo\thanks{{\LMF} \texttt{sylvie.boldo@inria.fr}}
  \and François Clément\thanks{{\SERENA} \texttt{francois.clement@inria.fr}}
  \and Florian Faissole\thanks{{\LMF} \texttt{F.Faissole@fr.merce.mee.com}}
  \and Vincent Martin\thanks{{\LMAC} \texttt{vincent.martin@utc.fr}}
  \and Micaela Mayero\thanks{{\LIPN} \texttt{mayero@lipn.univ-paris13.fr}}
}
\authorhead{S.~Boldo, F.~Clément, F.~Faissole, V.~Martin, \& M.~Mayero}

\RRtitle{\Titre}
\RRetitle{\Title}
\titlehead{\Title}

\RRresume{\Resume}
\RRabstract{\Abstract}

\RRmotcle{\Motscles}
\RRkeyword{\Keywords}

\RRprojets{Toccata et Serena}

\RCSaclay

\begin{document}

\RRNo{9401}
\makeRR

\clearpage
\section{Introduction}
\label{sec:intro}

This paper is dedicated to the {\Coq}~\cite{Link_Coq_ref} formalization of
Lebesgue integration theory.  Among many applications in mathematics, we focus
on the objective of building Sobolev spaces~\cite{ada:ss:75} that are used in
numerous fields: in functional analysis~\cite{yos:fa:80,bre:af:83,rud:rca:87},
and in statistical and probabilistic mathematics
\cite{tsy:ine:09,gv:fnb:17,fel:ipt:68,bil:pm:95,dur:pte:19}, to name just a
few.

Our main application is on the numerical resolution of Partial Differential
Equations (PDEs), using the {\FEM} (FEM).  Our final and long-term goal is to
formally prove the correctness of the FEM and of parts of a library
implementing it.  The FEM can be applied to compute numerical approximations to
solutions of many problems arising in physics, mechanics, and biology, for just
a few examples.  The success of the FEM is in large part due to its sound
mathematical foundation, see for instance
\cite{ztz:fem:13,cia:fem:02,qv:nap:94,eg:tpf:04} among the extensive
literature.  Prior to this work, we established in~\cite{BCF17} a formalization
of the proof of the {\LaxMilgramTh}, that is a relatively simple way of proving
the existence and uniqueness of the PDE solution and their FEM approximations
for a wide range of problems.  The {\LaxMilgramTh} is set on a general Hilbert
space (a complete vector space with an inner product).  In the context of PDEs,
the next stage is then the application of the {\LaxMilgramTh}: typically, for
the Poisson equation, one takes as Hilbert space a subspace of the~$H^1$
Sobolev space, see for instance \cite[Sec~3.2]{eg:tpf:04}.  The~$L^p$ Lebesgue
space is the space of functions whose absolute value to the power $p\geq1$ is
integrable, and~$H^1$ is defined as functions in~$L^2$ having a weak derivative
also in~$L^2$.  We recall that~$L^p$ is a Banach space (a complete normed
vector space), and~$L^2$ and~$H^1$ are Hilbert spaces.  This paper deals with
the construction of the Lebesgue integral for {\nonnegative} measurable
functions, a first step toward the formalization of~$L^p$, $H^1$, and other
Sobolev spaces.  Future work will include the formal definition and the proof
that they are indeed complete normed vector spaces.

As far as the integral is concerned, several options are available, {\eg}
see~\cite{bur:gi:07}.  The choice must be driven by the properties required for
our future developments.  As mentioned before, we are more interested in the
completeness of the considered functional spaces (like~$L^p$), than in the
ability to integrate the most exotic irregular functions.  On the one hand, the
Riemann integral is thus clearly not satisfactory as it is not compatible with
limit: the limit of Riemann-integrable functions is not necessarily a
Riemann-integrable function.  On the other hand, the gauge (\HK)
integral~\cite{kur:god:57,hen:ti:63,bar:mti:01} has attractive properties,
{\eg}, it is often considered as the easiest powerful integral to teach.
Unfortunately, its main drawback is that defining a complete normed vector
space of HK-integrable functions is not as obvious as with the Lebesgue~$L^p$
spaces~\cite{gw:faf:16,mt:dch:19}.  This led us to choose the Lebesgue
integral, which has the additional desirable property of being very general: it
is neither limited to functions defined on Euclidean spaces, nor to the
Lebesgue measure on~$\R^n$.

There are also several ways to build the integral of real-valued, or
complex-valued, functions for the Lebesgue measure.  First, the Daniell
approach~\cite{dan:gfi:18,fol:ram:99} allows the extension of an
\emph{elementary} integral defined for \emph{elementary} functions to a larger
class of functions by means of continuity and linearity.  When applied to the
Riemann integral for continuous real-valued functions with compact support, it
yields an integral equivalent to the Lebesgue integral for the Lebesgue
measure.  Second, a not so different alternate path consists in the completion
of the normed vector space of continuous functions with compact support, and
the extension of the Riemann integral which is uniformly
continuous~\cite{bou:int:65,die:ea2:68}.  Third, and the option we chose to
follow, is a modern form of the original works of Lebesgue~\cite{leb:lir:04}.
The Riemann integral is based on subdivisions of the domain of the function to
integrate.  In contrast, the Lebesgue approach focuses on the codomain.  For
each preimage, we need to provide its \emph{measure}, whatever its
irregularity.

This article covers the main concepts of measure theory such as the definitions
of $\sigma$-algebra, measurability of functions, measure, and simple functions.
Then, the integral is built following the Lebesgue scheme: first for
{\nonnegative} simple functions, then extended to all {\nonnegative} measurable
functions by taking the supremum.  The definition of the integral of a function
with arbitrary sign can be made by the difference, when possible, of the
integrals of the positive and negative parts of the function; this is out of
the scope of this paper and will be tackled in future~work.  The objective of
this paper is to formally prove the main results on {\nonnegative} measurable
functions: the {\BeppoLeviMonotConvTh}, and {\FatouLem}.

From a mathematical point of view, given a measure space defined by a set~$X$,
a $\sigma$-algebra~$\Sigma$, and a measure~$\mu$, the two statements can be
expressed in a mathematical setting as follows.

\begin{TextbookTheorem}[{\BL}, monotone convergence]
  \label{th:beppo-levi}
  \mbox{}\\
  Let $(f_n)_{n\in\N}$ be a sequence of {\nonnegative} measurable functions
  that is pointwise {\nondecreasing}.  Then, the pointwise limit
  $\lim_{n\to\infty}f_n$ is {\nonnegative} and measurable, and we have
  in~$\Rbarplus$
  \begin{equation}
    \label{eq:beppo-levi}
    \int \lim_{n \to \infty} f_n \, d\mu
    = \lim_{n \to \infty} \int f_n \, d\mu.
  \end{equation}
\end{TextbookTheorem}

\begin{TextbookTheorem}[{\Fl}]
  \label{th:fatou}
  \mbox{}\\
  Let $(f_n)_{n\in\N}$ be a sequence of {\nonnegative} measurable functions.
  Then, the pointwise limit $\liminf_{n\to\infty}f_n$ is {\nonnegative} and
  measurable, and we have in~$\Rbarplus$
  \begin{equation}
    \label{eq:fatou}
    \int \liminf_{n \to \infty} f_n \, d\mu
    \leq \liminf_{n \to \infty} \int f_n \, d\mu.
  \end{equation}
\end{TextbookTheorem}

These are the cornerstones of our intended future work, such as the building of
the~$L^p$ Lebesgue spaces as Banach spaces, the proofs of
{\LebesgueDominatedConvergenceTh} and of the {\TonelliFubiniThs}, and also the
construction of the Lebesgue measure (for instance through
{\CaratheodoryTh}~\cite{car:atm:63,dur:pte:19}).  As a consequence, we do not
yet need technical results on subset systems such as the
{\DynkinPiLambdaTh}~\cite{dur:pte:19}, or the {\MonotClassTh}~\cite{coh:mt:13},
that are popular tools for the extension of some property to the whole
$\sigma$-algebra ({\eg} the uniqueness of a measure).

\myskip

Interactive theorem proving is more and more being used and adapted for
formalizing real and numerical analysis.  Real-life applications, such as
hybrid systems or cyber-physical systems are critical and rely on advanced
analysis results.  Until now, only the Riemann integral was available
in~{\Coq}.  As useful as the Riemann integral is, the Lebesgue integral is
necessary for the numerical analysis we are examining.  In addition, even
though the Lebesgue integral exists in other theorem provers (see
Section~\ref{sec:soa}), we have decided to formalize it in {\Coq}.  Indeed, it
is crucial for our future work to be able to merge results both from numerical
analysis and from computer arithmetic (to bound rounding errors for instance).
For that, we plan to rely on the {\Flocq} library, which does not have a
comparable equivalent in other theorem provers.

We use the {\Coquelicot} library~\cite{BLM15}, a modernization of the real
standard library of {\Coq}, including a formalization of~$\Rbar$, described in
more detail in Section~\ref{sec:coquelicotRbar}.  This library provides
classical real numbers which correspond to the real analysis we deal with.  For
this reason, we have also decided, as basic choices of our formalization, to
use classical logic and to rely on the following axioms: strong excluded middle
and functional extensionality.  These choices are described in
Section~\ref{sec:basiccoq} and discussed in Section~\ref{sec:disc}.

The mathematical definitions and proofs were mainly taken from textbooks
\cite{mai:m2:14,gh:mip:13,gos:cms2:93}, detailed and compiled in a research
report~\cite{cm:li:21} in order to ease the formalization in {\Coq}.  The
{\Coq} code is available at
\mydisplay{\url{http://lipn.univ-paris13.fr/MILC/CoqLIntp/index.php},} or in
the public repository
\mydisplay{\url{https://lipn.univ-paris13.fr/coq-num-analysis/tree/LInt_p.1.0/Lebesgue},}
where the tag \texttt{LInt\_p.1.0} corresponds to the code of this article.

\myskip

The paper is organized as follows.  Section~\ref{sec:coqlibs} presents the main
basic {\Coq} choices on which our formalization is based.  The sequel is our
own contribution.  Section~\ref{sec:moreR} details auxiliary results on~reals.
The concept of measurability is discussed in Section~\ref{sec:measurable}, and
that of measure in Section~\ref{sec:measures}.  Section~\ref{sec:simple_fun} is
devoted to simple functions, and Section~\ref{sec:lint_p} to integration of
{\nonnegative} functions and the main theorems.  The case of the Dirac measure
is studied in Section~\ref{sec:Dirac}.  Concerns about proof engineering are
discussed in Section~\ref{sec:disc}.  Section~\ref{sec:soa} presents some state
of the art of the formalization of the integral.  Section~\ref{sec:concl}
concludes and gives some perspectives.

\section{{\Coquelicot} library and other basic {\Coq} choices}
\label{sec:coqlibs}

We first briefly review the few proof packages used in this work, and some
technical and logical choices we made.  These are discussed further in
Sections~\ref{sec:discuss:rbar} and~\ref{sec:discuss:logic}.

\subsection{The {\Coquelicot} library and~$\Rbar$}
\label{sec:coquelicotRbar}

The {\Coquelicot} library is a conservative extension of the {\Coq} real
standard library ({\Reals}), with total functions for limit, derivative, and
Riemann integral~\cite{BLM15,Lelay15coq,Lelay15}.  The features used here are
the generic topology, the hierarchy of algebraic structures based on canonical
structures, and the extended real numbers.

\paragraph{Generic topology.}
The {\Coquelicot} topology is defined using filters~\cite{car:tf:37,bou:tg:71}.
Intuitively, filters can be seen as sets of neighborhoods.  For instance, the
filter \coqe{eventually} on type \coqe{nat} corresponds to the most intuitive
neighborhoods of~$\infty$.
\begin{lstlisting}
Definition eventually : (nat -> Prop) -> Prop := fun P => \exists N, \forall n, N <== n -> P n.
\end{lstlisting}
It is used to define the convergence of sequences.

\paragraph{Algebraic hierarchy.}
{\Coquelicot} also defines an algebraic hierarchy based on canonical
structures.  A useful level here is \coqe{UniformSpace}, that formalizes the
mathematical concept of uniform space~\cite{wei:esu:37,bou:tg:71}: it is a
generalization of metric space with an abstraction of balls.  In a uniform
space~\coqe{E}, the property \coqe{open : (E -> Prop) -> Prop} characterizes
its open subsets.

\paragraph{Extended real numbers.}
{\Coquelicot} provides a definition of the extended real numbers
$\Rbar\myequal\R\cup\{-\infty,\infty\}$.  The formal definition contains three
constructors: \coqe{Finite} for real numbers, \coqe{p_infty} for~$\infty$ and
\coqe{m_infty} for~$-\infty$.  Conversely, the function \coqe{real} returns the
real number for finite numbers and~0 for~$\pm\infty$.
\begin{lstlisting}
Inductive Rbar :=
  | Finite : R -> Rbar
  | p_infty : Rbar
  | m_infty : Rbar.

Definition real : Rbar -> R :=
  fun x => match x with
    | Finite r => r
    | _ => 0
    end.
\end{lstlisting}

In addition to this definition, coercions from~\coqe{R} to~\coqe{Rbar} and
{\viceversa}, an order with \coqe{Rbar_lt} and \coqe{Rbar_le}, total operations
such as \coqe{Rbar_opp}, \coqe{Rbar_plus}, \coqe{Rbar_minus}, \coqe{Rbar_inv},
\coqe{Rbar_mult}, \coqe{Rbar_min} and \coqe{Rbar_abs} with their properties are
provided.

In particular, this means that addition on~$\Rbar$ is a total
function~\cite{BLM15} that always returns a value.  For instance,
$\infty+(-\infty)$ ({\ie} $\infty-\infty$) is~0, making some statements
unintuitive, see also Section~\ref{sec:meas_fn}.  However, the case of
multiplication is not an issue as the convention
$0\times\pm\infty=\pm\infty\times0=0$ is widely adopted for measure theory and
Lebesgue integration, because it yields more compact statements.

\subsection{Axioms}
\label{sec:basiccoq}

Real analysis, as most mathematics, uses classical logic, and measure theory
and Lebesgue integration are no exception.  For this reason, we chose to
conduct this formalization in a full-flavored classical framework.

We did not add our own axioms.  In addition to the axioms defining~\coqe{R}, we
require some classical properties from the standard library, listed here with
the theorems we use.

\begin{lstlisting}
Require Import ClassicalDescription.
Require Import PropExtensionality.
Require Import FunctionalExtensionality.
Require Import ClassicalChoice.

Check excluded_middle_informative.
        : \forall (P : Prop), {P} + {~p P}

Check propositional_extensionality.
        : \forall (P Q : Prop), P <-> Q -> P = Q

Check functional_extensionality.
        : \forall {A B : Type} (f g : A -> B), (\forallp x, f x = g x) -> f = g

Check choice.
        : \forall (A B : Type) (R : A -> B -> Prop),
            (\forallp x, \exists y, R x y) -> \exists (f : A -> B), \forall x, R x (f x)
\end{lstlisting}

We rely on \coqe{excluded_middle_informative} many times, including for
instance the definition of the characteristic function in
Section~\ref{sec:charac}.  We have a brief use of dependent types in
Section~\ref{sec:lint_simple} related to simple functions, and we then rely on
\coqe{propositional_extensionality} and \coqe{functional_extensionality}.
Last, we rely on \coqe{choice} at a single point in the proof of Lemma
\coqe{negligible_union_countable}, and this is explained in
Section~\ref{sec:measures:negl}.

\section{Auxiliary results about the reals}
\label{sec:moreR}

From now on, we present our own contributions.  A global dependency graph of
our {\Coq} files and results is given in Figure~\ref{fig:dep}
page~\pageref{fig:dep}, with links back to the appropriate sections.

The theorems described in this section are not dedicated to the Lebesgue
integral and could be part of a support library.  In Section~\ref{sec:topoR},
we show the expression of open subsets of~$\R$ and~$\R^2$ with a countable
topological basis.  Section~\ref{sec:sum_Rbar} deals with sums on~$\Rbar$.
Section~\ref{sec:limit} presents some additional results about limits.

\subsection{Second-countability of real numbers}
\label{sec:topoR}

In Section~\ref{sec:gen_equiv}, we need to characterize and decompose the open
subsets of~$\R$.  More precisely, we build generators of the $\sigma$-algebras
of~$\R$ and~$\R^2$ that contain the open subsets.  Such generators need to be
of countable size to comply with the properties of $\sigma$-algebras.  Thus,
the concepts of topological basis and second-countability appeared necessary.

Recall that a topological basis allows to express any open subset of a
topological space as the union of a subfamily of the basis.  A topological
space is called \emph{second-countable} when it admits a countable topological
basis.  Euclidean spaces~$\R^n$ are second-countable.  Indeed, the open boxes
with rational boundaries form such a countable topological basis.  In the case
of~$\R$, this is expressed as follows.  For any open subset~$A$ of~$\R$, there
exists a sequence of pairs of rationals $(q_1^n,q_2^n)_{n\in\N}$ such that~$A$
is the union of the corresponding open intervals,
$A=\bigcup_{n\in\N}(q_1^n,q_2^n)$.

The mathematical proof is well known, but the road to formalization was
tedious.

\paragraph{Countability.}
We define bijections from~$\N$ to~$\N^2$, $\Z$, $\Q$ and~$\Q^2$.  It is not
enough to prove they have the same size, we need ``perfect'' bijections,
meaning inverse functions from one type to the other, handling correctly
special cases such as zero.

\paragraph{Connected components.}
Given a subset~$A$ of~$\R$ and a real~$x$, we define the bounds of the largest
possible interval included in~$A$ and containing~$x$ ({\aka} the connected
component of~$x$ in~$A$).
\begin{lstlisting}
Definition bottom_interv : (R -> Prop) -> R -> Rbar :=
  fun A x => Glb_Rbar (fun z => \forall y, z < y < x -> A y).

Definition top_interv : (R -> Prop) -> R -> Rbar :=
  fun A x => Lub_Rbar (fun z => \forall y, x < y < z -> A y).
\end{lstlisting}
The functions \coqe{Glb_Rbar} and \coqe{Lub_Rbar} are total functions from the
{\Coquelicot} library that compute the greatest lower bound and the least upper
bound of a subset of reals.

We prove many properties such as belonging to the closure of its own connected
component, \coqe{bottom_interv A x <== x <== top_interv A x}, and belonging to
the interior of its own connected component for points of open subsets, that is
\coqe{open A -> A x ->}\linebreak[0]%
\coqe{bottom_interv A x < x < top_interv A x}.

\paragraph{Using density of rational numbers.}
Given an open subset~$A$ of~$\R$, we prove through density of~$\Q$ in~$\R$
that~$A$ contains at most a countable number of connected components.
\begin{lstlisting}
Lemma open_R_charac_Q :
  \forall (A : R -> Prop), open A ->
    \forall x, A x <-> (\existsp q : Q, let y := Q2R q in
      A y /\ Rbar_lt (bottom_interv A y) x /\ Rbar_lt x (top_interv A y)).
\end{lstlisting}
Countability appears in this lemma through the rationality of~$y$ (otherwise,
the theorem would be trivial by using~$x$).

In addition, using again the density of~$\Q$ in~$\R$, we can take rational
bounds for these intervals (by taking countable unions of intervals with
rational bounds to recover each initial interval with real bounds).  And then,
using countability of~$\Q^2$, we have a bijection from the integers to the
rational bounds of the open intervals, and these serve as topological basis.
\begin{lstlisting}
Definition topo_basis_R : nat -> R -> Prop :=
  fun n x => Q2R (fst (bij_NQ2 n)) < x < Q2R (snd (bij_NQ2 n)).
\end{lstlisting}

\paragraph{Second-countability.}
Given an open subset~$A$ of~$\R$, we want to exhibit the~$q_i^n$'s such that
$A=\bigcup_{n\in\N}(q_1^n,q_2^n)$.  This means we need to choose among the
possible intervals of the topological basis the useful ones by relying on a
property~$P$.  Then, $A$~is equivalent to the countable union of the
\coqe{topo_basis_R n} such that~\coqe{P n} holds.
\begin{lstlisting}
Lemma R_second_countable :
  \forall (A : R -> Prop), open A ->
    \exists (P : nat -> Prop), (\forallp x, A x <-> \exists n, P n /\ topo_basis_R n x).
\end{lstlisting}

The same property holds for~$\R^2$.  We can define a topological basis
for~$\R^2$ (from the tensor product of the topological basis of~$\R$) and prove
\begin{lstlisting}
Lemma R2_second_countable :
  \forall (A : R * R -> Prop), open A ->
    \exists (P : nat -> Prop), (\forallp x, A x <-> \exists n, P n /\ topo_basis_R2 n x).
\end{lstlisting}

\subsection{About sums of extended real numbers}
\label{sec:sum_Rbar}

Integrals of simple functions are defined in Section~\ref{sec:lint_simple} as
sums of extended reals.  Even if we only sum {\nonnegative} extended reals, we
decided to use only \coqe{Rbar} as discussed in Section~\ref{sec:discuss:rbar}.
But as in mathematics, the addition on~\coqe{Rbar} as defined by {\Coquelicot}
is not associative.  Indeed, $\infty+(\infty-\infty)=\infty$, while
$(\infty+\infty)-\infty=0$.  Our design choice therefore implies that big
operators~\cite{Bertot2008canonical} cannot be used.

\myskip

Let us begin with sums of a finite number of values.  The definition goes as
expected, with an equivalent alternative using \coqe{fold_right} for lists
instead of functions.
\begin{lstlisting}
Fixpoint sum_Rbar n (f : nat -> Rbar) : Rbar :=
  match n with
  | 0 => f 0%nat
  | S n1 => Rbar_plus (f (S n1)) (sum_Rbar n1 f)
  end.

Definition sum_Rbar_l : list Rbar -> Rbar := fun l => fold_right Rbar_plus 0 l.
\end{lstlisting}

In addition, we found it useful to define an ``applied'' sum that takes a
function~$f$ and a list~$\ell$ and returns the sum of the images
$\Sigma_{i\in\ell}f(i)$.
\begin{lstlisting}
Definition sum_Rbar_map : \forall {E : Type}, list E -> (E -> Rbar) -> Rbar :=
  fun E l f => sum_Rbar_l (map f l).
\end{lstlisting}
The curly brackets around~\coqe{E} mean that this argument is implicit and need
not be specified, as {\Coq} can guess it from the type of the list~\coqe{l}.

This definition allows us to use extensionality either on the list~\coqe{l}, on
the function~\coqe{f}, or on the application \coqe{map f l}, which turned out
to be more practical than what this obvious definition seems.  Examples of use
are the following lemmas (that do not need {\nonnegativity}).  The first one
mixes two applications.
\begin{lstlisting}
Lemma sum_Rbar_map_map :
  \forall {E F : Type} (f : E -> F) (g : F -> Rbar) (l : list E),
    sum_Rbar_map (map f l) g = sum_Rbar_map l (fun x => g (f x)).
\end{lstlisting}

The second one focuses on the statement
$\Sigma_{i\in\ell_1}f(i)=\Sigma_{i\in\ell_2}f(i)$.  Such result is obvious when
$\ell_1=\ell_2$, but it is also possible to prove it when the lists are
identical except for items~$i$ of the lists such that $f(i)=0$, as these do not
impact the final sums.  Indeed, the sums may be the same even if the two lists
are different (and of different lengths for instance).  The function
\coqe{select} is defined later in Section~\ref{sec:canonizer}.  It has type
\coqe{(E -> Prop) -> list E -> list E} and selects the elements of a list that
have a given property, without changing otherwise the order in the lists.
\begin{lstlisting}
Lemma sum_Rbar_map_ext_l :
  \forall {E : Type} (l1 l2 : list E) (f : E -> Rbar),
    select (fun x => (f x <> 0)) l1 = select (fun x => f x <> 0) l2 ->
    sum_Rbar_map l1 f = sum_Rbar_map l2 f.
\end{lstlisting}

\myskip

When values are {\nonnegative}, associativity is back and we have the expected
theorems on~sums.
\begin{lstlisting}
Lemma sum_Rbar_end :
  \forall f n, (\forallp i, (i <== S n)%nat -> Rbar_le 0 (f i)) ->
    (sum_Rbar (S n) f = Rbar_plus (f 0%nat) (sum_Rbar n (fun i => f (S i)))).

Lemma sum_Rbar_l_concat :
  \forall (l1 l2 : list Rbar), non_neg_l l1 -> non_neg_l l2 ->
    sum_Rbar_l (l1 ++ l2) = Rbar_plus (sum_Rbar_l l1) (sum_Rbar_l l2).
\end{lstlisting}
For the sake of brevity, we have defined the properties \coqe{non_neg} and
\coqe{non_neg_l} for {\nonnegative} functions and lists.

The most interesting theorem about sums of lists is the ability to swap the
order of a double summation,
\[
  \sum_{i_1 \in \ell_1} \sum_{i_2 \in \ell_2} f (i_1, i_2)
  = \sum_{i_2 \in \ell_2} \sum_{i_1 \in \ell_1} f (i_1, i_2).
\]
\begin{lstlisting}
Lemma sum_Rbar_map_switch :
  \forall {E : Type} (f : E -> E -> Rbar) l1 l2,
    (\forallp x y, In x l1 -> In y l2 -> Rbar_le 0 (f x y)) ->
    sum_Rbar_map l1 (fun x => sum_Rbar_map l2 (fun y => f x y)) =
      sum_Rbar_map l2 (fun y => sum_Rbar_map l1 (fun x => f x y)).
\end{lstlisting}

\subsection{About limits}
\label{sec:limit}

We also need some additional results on limits and suprema.

First of all, the sums defined in Section~\ref{sec:sum_Rbar} have a finite
number of terms.  But the main theorems to come rely on infinite sums ({\ie}
series).  The most common definition is the limit of the finite partial sums,
{\ie} \coqe{Lim_seq} in {\Coquelicot} \cite{BLM15}.  Nevertheless, by virtue of
the least-upper-bound property in~$\R$ and~$\Rbar$, when a sequence is
increasing (which happens when adding only {\nonnegative} values), the supremum
is also the limit, and we may equivalently use \coqe{Sup_seq} instead.  This
has proved more convenient and more suited to our needs.  So theorems of
Section~\ref{sec:lint_p} such as the {\BeppoLeviTh} rely on \coqe{Sup_seq}.

Next, we are interested in the limit inferior of sequences in~$\Rbar$.  But,
{\Coquelicot} only provides \coqe{LimInf_seq} of type
\coqe{(nat -> R) -> Rbar}, and nothing for \coqe{nat -> Rbar} sequences.
Therefore, we defined a minor variant of the desired type, and proved a few
lemmas by directly copying their proofs from those for \coqe{LimInf_seq} in
{\Coquelicot}.
\begin{lstlisting}
Definition LimInf_seq' : (nat -> Rbar) -> Rbar :=
  fun u => Sup_seq (fun m => Inf_seq (fun n => u (n + m)%nat)).
\end{lstlisting}

\section{Measurability}
\label{sec:measurable}

We present now the formalization of $\sigma$-algebras, which are defined as an
inductive type.  They~characterize \emph{measurable} subsets, and particular
attention is paid to~$\R$, $\Rbar$ and~$\R^2$, where the open subsets generate
the \emph{Borel} measurable subsets.

The issue of subsets is briefly addressed in Section~\ref{sec:charac}.
Section~\ref{sec:sigma-alg} is devoted to the measurability of subsets, and
Section~\ref{sec:cartesian_measu} to Cartesian products.  The Borel subsets
of~$\R$ and~$\Rbar$ are detailed in Section~\ref{sec:gen_equiv}.  And
Section~\ref{sec:meas_fn} deals with the measurability of functions.

\subsection{Subsets and characteristic functions}
\label{sec:charac}

We consider a generic set~\coqe{E} defined in {\Coq} as \coqe{E : Type}.
Usually, subsets of~\coqe{E} are defined in {\Coq} as having type
\coqe{E -> Prop}, or \coqe{E -> bool}.  We choose \coqe{Prop}, and this is
discussed in Section~\ref{sec:discuss:logic}.  Then, the power set of~\coqe{E}
has type \coqe{(E -> Prop) -> Prop}.

Given a subset~$A$, we define its characteristic function (or indicator
function)~$\charac{A}$ that maps elements of~$A$ to~1, and others to~0.
\begin{lstlisting}
Context {E : Type}.

Definition charac : (E -> Prop) -> E -> R :=
  fun A x => match (excluded_middle_informative (A x)) with
    | left _ => 1
    | right _ => 0
    end.
\end{lstlisting}
Indeed, it is very convenient for direct use in arithmetic expressions without
exhibiting the membership conditional in a dependent type or an assumption.  It
is used a lot in the context of simple functions in
Section~\ref{sec:simple_fun}.

The characteristic function is also convenient to simulate the restriction of a
numerical function to a subset, for instance in Section~\ref{sec:meas_fn}.
More precisely, the mathematical function~$\restr{f}{A}$ could be formalized
either as a record with a dependent type, or as a total function.  We have
explored the first way which became impractical as proofs creep into our
statements and prevent some rewritings.  The total function is then
$f\times\charac{A}$, which is the correct value when needed and~0 elsewhere.
This is perfectly suited to our context, as integrating zero has no impact.

\subsection{Measurability of subsets}
\label{sec:sigma-alg}

The design choice for the measurability of subsets, {\ie} the definition of
$\sigma$-algebra, is a central issue for this paper.  Even though several
equivalent definitions are possible, the use of an inductive type has proved
successful, with several proofs done by induction.

Among several possible informal definitions~\cite[Section~8.6]{cm:li:21}, a
$\sigma$-algebra is a subset of the power set that contains the empty set, and
is closed under complement and countable unions.  In fact, a $\sigma$-algebra
can be really huge and it is very convenient to represent it with a smaller
collection~$G$ of so-called generators, and to consider the smallest
$\sigma$-algebra containing~$G$.  This corresponds to the informal concept of
\emph{generated} $\sigma$-algebra.  Indeed, in many situations, it is
sufficient to establish a property on~$G$ to have it on the whole
$\sigma$-algebra generated by~$G$.

While this may suggest the use of a record, we rely on an inductive type.  More
precisely, we ``start'' with a collection of generators
\coqe{genE : (E -> Prop) -> Prop}.  Then, a subset is measurable if it is
either a generator, empty, the complement of a measurable subset, or the
countable union of measurable subsets.  This design choice is discussed in
Section~\ref{sec:discuss:measurability}.  Note that the issue of generators is
at the center of Section~\ref{sec:cartesian_measu} for Cartesian products, and
discussed in Section~\ref{sec:gen_equiv} for the Borel subsets of real numbers.
\begin{lstlisting}
Variable genE : (E -> Prop) -> Prop.

Inductive measurable : (E -> Prop) -> Prop :=
  | measurable_gen : \forall A, genE A -> measurable A
  | measurable_empty :  measurable (fun _ => False)
  | measurable_compl : \forall A, measurable (fun x => ~ A x) -> measurable A
  | measurable_union_countable :
      \forall (A : nat -> E -> Prop), (\forallp n, measurable (A n)) -> measurable (fun x => \exists n, A n x).
\end{lstlisting}

From this definition, we then prove various lemmas, relying on our classical
setting, such as measurability of the full set, and of countable intersections.
\begin{lstlisting}
Lemma measurable_inter_countable :
  \forall (A : nat -> E -> Prop), (\forallp n, measurable (A n)) -> measurable (fun x => \forall n, A n x).
\end{lstlisting}
A mathematically unexpected, but quite useful theorem is the following.
\begin{lstlisting}
Lemma measurable_Prop : \forall P, measurable (fun _ => P).
\end{lstlisting}
Constant properties (that do not depend on a variable), be they true or false,
are measurable as both \coqe{True} (the full set) and \coqe{False} (the empty
set) are measurable.  When decomposing a subset to prove its measurability,
this comes in handy.

\myskip

In many situations, several collections of generators are possible, and
switching between them may be convenient for the proof at hand.  In fact,
if~$G_1$ is included in the $\sigma$-algebra generated by~$G_2$, and
{\viceversa}, then both generated $\sigma$-algebras are the same.  This yields
the following extensionality result.
\begin{lstlisting}
Lemma measurable_gen_ext :
  \forall genE1 genE2,
    (\forallp A, genE1 A -> measurable genE2 A) -> (\forallp A, genE2 A -> measurable genE1 A) ->
    (\forallp A, measurable genE1 A <-> measurable genE2 A).
\end{lstlisting}

\myskip

We now define what is a $\sigma$-algebra, but this definition is hardly used
later on as we rely mostly on the previous inductive.  A $\sigma$-algebra is
formally defined as a subset of the power set that is equal to the
$\sigma$-algebra induced by itself as generator.
\begin{lstlisting}
Definition is_sigma_algebra: ((E -> Prop) -> Prop) -> Prop :=
  fun calS => calS = measurable calS.
\end{lstlisting}
We have the equivalence with one of the commonly used mathematical definitions:
$\calS$~is a $\sigma$-algebra when it contains the empty set, and is closed
under complement and countable unions.
\begin{lstlisting}
Lemma is_sigma_algebra_correct :
  \forall calS, is_sigma_algebra calS <->
    (calS (fun _ => False) /\
    (\forallp A, calS (fun x => ~ A x) -> calS A) /\
    (\forallp (A : nat -> E -> Prop), (\forallp n, calS (A n)) -> calS (fun x => \exists n, A n x))).
\end{lstlisting}
We can of course prove that the basic $\sigma$-algebras are indeed compliant
with our definition, be it the discrete $\sigma$-algebra (the whole power set),
or the trivial $\sigma$-algebra (reduced to $\{\emptyset,E\}$).
\begin{lstlisting}
Lemma is_sigma_algebra_discrete : is_sigma_algebra (fun _ => True).
Lemma is_sigma_algebra_trivial : is_sigma_algebra (fun A => (\forallp x, ~ A x) \/ (\forallp x, A x)).
\end{lstlisting}

An immediate consequence of the extensionality result about generators is the
idempotence of $\sigma$-algebra generation.  Indeed, the $\sigma$-algebra
generated by a given generated $\sigma$-algebra is the very same
$\sigma$-algebra.  This may be expressed in {\Coq} as
\coqe{is_sigma_algebra (measurable genE)}, showing that our definition is
indeed a $\sigma$-algebra in the mathematical sense.  In addition, the
definition by induction gives us for free that our definition represents the
smallest generated $\sigma$-algebra.

\myskip

To sum up, in our development, the measurability of subsets of \coqe{E : Type}
is built by induction from a \emph{generator}
\coqe{genE : (E -> Prop) -> Prop}, providing a $\sigma$-algebra.

\subsection{Cartesian product and measurability}
\label{sec:cartesian_measu}

Although we do not deal with the {\TonelliFubiniThs} in this paper, the
Cartesian product is used in Section~\ref{sec:meas_fn} to establish
measurability of the addition and multiplication of two measurable numerical
functions.

Given two measurable spaces, {\ie} two sets~$E$ and~$F$ and their associated
generators~$G_E$ and~$G_F$, it is natural to ask the question of measurability
on the Cartesian product $E\times F$; but with which $\sigma$-algebra?  Among
other possibilities, the tensor product of the two $\sigma$-algebras is of
paramount interest, since it makes both canonical projections (the maps
$((x_E,x_F)\mapsto x_E)$ and $((x_E,x_F)\mapsto x_F)$) measurable.  It is the
$\sigma$-algebra generated by the Cartesian products of measurable subsets
of~$E$ and~$F$.

Unfortunately, on the matter of generator, simply taking the Cartesian products
of elements of~$G_E$ and~$G_F$ is not correct: in this case, for instance, one
cannot prove the measurability of $A_E \times F$, for $A_E\in G_E$.  We need to
add the full sets to the initial generator, using the following definition.
\begin{lstlisting}
Definition gen2 : (E * F -> Prop) -> Prop :=
  fun A => \exists AE AF, (genE AE \/ AE = fun _ => True) /\ (genF AF \/ AF = fun _ => True) /\
    (\forallp X, A X <-> AE (fst X) /\ AF (snd X)).
\end{lstlisting}
And we prove this satisfies the desired property.
\begin{lstlisting}
Lemma gen2_is_product_measurable :
  \forall AE AF, measurable genE AE -> measurable genF AF ->
    measurable (gen2 genE genF) (fun X => AE (fst X) /\ AF (snd X)).
\end{lstlisting}

\subsection{Borel subsets of real numbers}
\label{sec:gen_equiv}

We specify now an important class of $\sigma$-algebras.  When the measurable
space has also a topological space structure ({\eg} \coqe{UniformSpace} in
{\Coquelicot}, see Section~\ref{sec:coquelicotRbar}), one usually selects the
Borel $\sigma$-algebra.  It is generated by all the open subsets, or
equivalently by all the closed subsets, and has the nice property of providing
measurability for continuous functions (see Section~\ref{sec:meas_fn}).

Lebesgue integration theory is essentially meant for real-valued functions (or
with codomain~$\Rbar$, $\R^n$, or~$\C^n$).  Thus, we need to equip~$\R$,
and~$\Rbar$, with their Borel $\sigma$-algebras, and we have some leeway in
choosing the generators, instead of all open subsets.  Now, we present our
choice, and also prove that other possibilities define the same
$\sigma$-algebras.

\paragraph{Borel subsets on~$\R$.}
Among many possibilities, we pick the closed intervals (of the form $[a,b]$,
with $a\leq b$ reals) for~$\R$.
\begin{lstlisting}
Definition gen_R_cc : (R -> Prop) -> Prop := fun A => \exists a b, (\forallp x, A x <-> a <== x <== b).
\end{lstlisting}
This choice of \coqe{gen_R_cc} is somewhat arbitrary, and could be changed.
Thus, we introduce an anonymous \coqe{gen_R} that will be used in the sequel of
the paper for the definition of measurable subsets of~$\R$.
\begin{lstlisting}
Definition gen_R := gen_R_cc.
Definition measurable_R : (R -> Prop) -> Prop := measurable gen_R.
\end{lstlisting}

Other choices for~$\R$ include the open intervals, or of the form $[a,b)$, or
the open left-rays of the form $(-\infty,b)$.  We proved for instance that
measurability on~$\R$ generated by closed intervals (our definition) is the
same as measurability generated by open intervals.
\begin{lstlisting}
Definition gen_R_oo : (R -> Prop) -> Prop := fun A => \exists a b, \forall x, A x <-> a < x < b.
Lemma measurable_R_equiv_oo : \forall A, measurable_R A <-> measurable gen_R_oo A.
\end{lstlisting}
The proof is a call to the generator extensionality lemma, and then relies on
basic properties of measurability (closedness under complement and countable
union), and on the definition of a nested sequence of closed intervals (from
\coqe{gen_R_cc}) whose union is an open interval (from \coqe{gen_R_oo}), thanks
to the Archimedean property of~$\R$.  Moreover, from the density of~$\Q$
in~$\R$, we may only consider open intervals with rational endpoints.
\begin{lstlisting}
Definition gen_R_Qoo : (R -> Prop) -> Prop := fun A => \exists a b, \forall x, A x <-> Q2R a < x < Q2R b.
Lemma measurable_R_equiv_Qoo : \forall A, measurable_R A <-> measurable gen_R_Qoo A.
\end{lstlisting}

And finally, more interestingly from a mathematical viewpoint, we prove that
our measurable subsets on~$\R$ (based on closed intervals) are indeed the Borel
subsets generated by \coqe{open} from \coqe{UniformSpace}.
\begin{lstlisting}
Lemma measurable_R_open : \forall A, measurable_R A <-> measurable open A.
\end{lstlisting}
The proof is simply an application of lemma \coqe{R_second_countable} from
Section~\ref{sec:topoR} stating that any open subset is the countable union of
open intervals with rational endpoints.  This is needed in
Section~\ref{sec:meas_fn} where the measurability of the addition of two
measurable real-valued functions relies on the continuity of the addition
in~$\R$.

\paragraph{Borel subsets on~$\R^2$.}
Combining the generator for a Cartesian product of
Section~\ref{sec:cartesian_measu} and the second-countability of~$\R^2$ of
Section~\ref{sec:topoR}, we have an equivalence result for the Borel subsets
of~$\R^2$.
\begin{lstlisting}
Definition gen_R2 : (R * R -> Prop) -> Prop := gen2 gen_R gen_R.
Definition measurable_R2 : (R * R -> Prop) -> Prop := measurable gen_R2.
Lemma measurable_R2_open : \forall (A : R * R -> Prop), measurable_R2 A <-> measurable open A.
\end{lstlisting}
Here, \coqe{open} stands for the open subsets of~$\R^2$.  The canonical
structures of {\Coquelicot} deduce that~$\R^2$, as product of two
\coqe{UniformSpace}s, is a \coqe{UniformSpace}.

\paragraph{Borel subsets on~$\Rbar$.}
For~$\Rbar$, the generators we choose are the closed right-rays (of the form
$[a,\infty]$, $a\in\Rbar$), but we also define an anonymous \coqe{gen_Rbar}.
\begin{lstlisting}
Definition gen_Rbar_cu : (Rbar -> Prop) -> Prop := fun A => \exists a, \forall x, A x <-> Rbar_le a x.
Definition gen_Rbar := gen_Rbar_cu.
Definition measurable_Rbar : (Rbar -> Prop) -> Prop := measurable gen_Rbar.
\end{lstlisting}
We proved the equivalence with the measurability defined by closed left-rays
(of the form $[-\infty,a]$).  Unlike~$\R$, the measurability of the addition of
two measurable $\Rbar$-valued functions does not rely on continuity anymore
(see Section~\ref{sec:meas_fn}), and we did not prove that our measurable
subsets on~$\Rbar$ (based on closed rays) are indeed the Borel subsets
generated by the open subsets of~$\Rbar$, as we do not need it for now.

Next, we proved that measurability is compatible with scaling.
\begin{lstlisting}
Lemma measurable_scal_Rbar :
  \forall A l, measurable_Rbar A -> measurable_Rbar (fun x => A (Rbar_mult l x)).
\end{lstlisting}
Note that~$\ell$ may be any extended real, even~0 or~$\pm\infty$.  So one may
imagine the numerous subcases to ensure this lemma.

\subsection{Measurability of functions}
\label{sec:meas_fn}

From the measurability of subsets defined above, we can now define the
measurability of a function.

\paragraph{General case.}
Given two sets~$E$ and~$F$ and associated generators~$G_E$ and~$G_F$, a
function $f:E\to F$ is measurable when for every measurable subset~$A$, the
subset~$f^{-1}(A)$ is measurable, {\ie} $\{x\,|\,f(x)\in A\}$ is measurable.
Note that~$f^{-1}$ is obviously understood as a function from the power set
of~$F$ to the one of~$E$.  The measurability is then defined in {\Coq} as
follows.
\begin{lstlisting}
Definition measurable_fun : (E -> F) -> Prop :=
  fun f => \forall A, measurable genF A -> measurable genE (fun x => A (f x)).
\end{lstlisting}

We then prove some basic properties.  For instance, it is enough to consider
the generators to ensure the measurability of a function.
\begin{lstlisting}
Lemma measurable_fun_gen :
  \forall (f : E -> F), measurable_fun f <-> (\forallp A, genF A -> measurable genE (fun x => A (f x))).
\end{lstlisting}

When~\coqe{E} and~\coqe{F} are also \coqe{UniformSpace} (from {\Coquelicot},
see Section~\ref{sec:coquelicotRbar}), the use of Borel $\sigma$-algebras
(generated by the open subsets) ensures that continuous functions are
measurable.  As~explained in Section~\ref{sec:coquelicotRbar}, the continuity
definition is based on filters.
\begin{lstlisting}
Lemma measurable_fun_continuous :
  \forall f, (\forallp x, continuous f x) -> measurable_fun open open f.
\end{lstlisting}
This is simply due to the fact that the inverse image of an open subset by a
continuous function is an open subset.

\paragraph{Case of numerical functions.}
Now let us consider the case of numerical functions, with codomain~$\R$,
or~$\Rbar$.  The definition relies on the generators \coqe{gen_R} and
\coqe{gen_Rbar} defined above.
\begin{lstlisting}
Definition measurable_fun_R : (E -> R) -> Prop := measurable_fun genE gen_R.
Definition measurable_fun_Rbar : (E -> Rbar) -> Prop := measurable_fun genE gen_Rbar.
\end{lstlisting}

Later on, we have to deal with piecewise-defined functions, and in such a
situation, it is interesting to treat each piece separately, and to use the
restriction defined in Section~\ref{sec:charac} as the multiplication by the
characteristic function.  The following result, simple but useful, states that
given a measurable subset~$A$ and a measurable function~$g$, given a
function~$f$ equal to~$g$ on~$A$, then $f\times\charac{A}$ is measurable.  Its
proof is rather easy given the proved properties of the measurability
of~subsets.
\begin{lstlisting}
Lemma measurable_fun_when_charac :
  \forall (f g : E -> Rbar) A, measurable genE A ->
    (\forallp x, A x -> f x = g x) -> measurable_fun_Rbar g ->
    measurable_fun_Rbar (fun x => Rbar_mult (f x) (charac A x)).
\end{lstlisting}

\myskip

The main mathematical result of the rest of this section is the compatibility
of measurability of functions with algebraic operations (addition, scalar
multiplication and multiplication); the most complex one being the addition.
From the mathematical standpoint, when extended real values are involved, it is
assumed that these operations are well-defined.  In {\Coq}, when using
operations on~\coqe{Rbar} from {\Coquelicot} that are total functions, the
situation is different, and somewhat more complex as explained below.

\paragraph{Functions to~$\R$.}
Let us prove first the measurability of the sum of two measurable
real-valued~functions.
\begin{lstlisting}
Lemma measurable_fun_Rplus :
  \forall f1 f2, measurable_fun_R f1 -> measurable_fun_R f2 ->
    measurable_fun_R (fun x => f1 x + f2 x).
\end{lstlisting}

The proof uses the compatibility of measurability with the composition of
functions: if both~$f$ and~$g$ are measurable, then so is $f\circ g$.  This is
applied to $f\eqdef((x,y)\mapsto x+y)$ of type $\R^2\to\R$ and
$g\eqdef(x\mapsto(f_1(x),f_2(x)))$ of type $\R\to\R^2$.

Measurability on~$\R^2$ relies on \coqe{gen_R2}, the generator of the Borel
subsets of~$\R^2$ defined in Section~\ref{sec:gen_equiv}.  The proof is based
on the generator equivalence between \coqe{gen_R2} and \coqe{open}, and on the
continuity of addition.  This proof was not difficult, but happened to be much
higher-level than~expected.  The multiplication of real-valued functions is
treated exactly in the same way.

Scalar multiplication for measurable functions is deduced from a similar
theorem about scalar multiplication for measurable subsets.  In the end, the
measurable real-valued functions form an algebra (over the field~$\R$); however
we have not stated it (with canonical structures for instance) as we have no
use for it, but all the needed lemmas are proved.

\paragraph{Functions to~$\Rbar$.}
Let us consider now the addition of measurable extended real-valued functions.
The semantics of~$+_{\Rbar}$ is more complex, as it raises the question of what
is $\infty-\infty$.  We rely on the {\Coquelicot} definition of
\coqe{Rbar_plus}.  As a total function, it returns~0 in this special case, see
Section~\ref{sec:coquelicotRbar}.  The proof for~$\R$ was based on the
continuity of~$+$; but that cannot be used here, as~$+_{\Rbar}$ is not
continuous on the whole set~$\Rbar^2$ (there are problems at infinity, even for
the total function).

In order to stick closely to the mathematics, we rely on a way to express the
legality of addition: the property \coqe{ex_Rbar_plus} that basically prevents
adding~$\infty$ and~$-\infty$.  Thus, we prove the following~theorem.
\begin{lstlisting}
Lemma measurable_fun_plus :
  \forall f1 f2, measurable_fun_Rbar f1 -> measurable_fun_Rbar f2 ->
    (\forallp x, ex_Rbar_plus (f1 x) (f2 x)) ->
      measurable_fun_Rbar (fun x => Rbar_plus (f1 x) (f2 x)).
\end{lstlisting}
The proof is a little tedious as it splits~$E$ into all the possible cases
using \coqe{measurable_}\linebreak[0]%
\coqe{fun_when_}\linebreak[0]%
\coqe{charac}: when both~$f_1(x)$ and~$f_2(x)$ are finite, the previous theorem
on~$\R$ is used.  Otherwise, the preimages of~$\pm\infty$ are measurable since
singletons are (as closed subsets).  Thus, we are able to finish all the cases.

\myskip

Among the peculiarities of {\Coq} compared to mathematics, note that a simpler
theorem can be~devised.
\begin{lstlisting}
Lemma measurable_fun_plus' :
  \forall f1 f2, measurable_fun_Rbar f1 -> measurable_fun_Rbar f2 ->
    measurable_fun_Rbar (fun x => Rbar_plus (f1 x) (f2 x)).
\end{lstlisting}
It states the same conclusion, but without assuming the legality of addition.
Indeed, the total function $(x\mapsto f_1(x)+_{\Rbar}f_2(x))$, with value~0
when both operands are infinite opposites, is actually measurable.  This
subtlety when considering $\infty-\infty$ is related to total functions, a
design choice that prevents dependent types but may give strange results when
out of the domain of the function.  This strangeness also exists in the {\Coq}
standard library of reals~\cite{May01} when considering the division as a total
function, making~\coqe{1/0} a valid real.  This hard question would be solved
more naturally in other provers, for instance in {\PVS} relying on TCCs
(Type-Correctness Conditions)~\cite{Owre1999pvs:alt}.  To~conclude, the main
problem with this theorem is that it does not state what the mathematicians
read in it, so we have decided not to use it.

The multiplication of two functions taking their values in~$\Rbar$ is treated
similarly.  However, it does not raise the same issues as addition, because
{\Coquelicot} and mathematics for measure theory use the same convention
$\pm\infty\times0=0$, see Section~\ref{sec:coquelicotRbar}.  Multiplication by
a scalar is deduced from a similar theorem on measurable subsets.  Note that in
contrast to the case of~$\R$, measurable functions with values in \coqe{Rbar}
do not form an algebra, as \coqe{Rbar_plus} is not associative, see
Section~\ref{sec:sum_Rbar}.

\section{Measure}
\label{sec:measures}

A measurable space with a $\sigma$-algebra can be equipped with a measure.  A
measure is a mapping from measurable subsets to {\nonnegative} extended real
values that satisfies additivity properties.  Some well-known measures are the
Lebesgue measure, the counting measure, the Dirac measure (see
Section~\ref{sec:Dirac}), and numerous probability measures (that take values
in the interval~$[0,1]$).

Measure theory is a general abstract setting that applies to any measure, and
the axiomatization of their fundamental properties is formalized here with an
instantiation in Section~\ref{sec:Dirac}.

\subsection{Specification and basic properties}
\label{sec:measures:spec_meas}

Given a measurable space defined by a set \coqe{E : Type} and a
generator~\coqe{genE} of type \coqe{(E -> Prop) -> Prop} (see
Section~\ref{sec:sigma-alg}), our design choice is to specify measures as a
\coqe{Record} type containing a map \coqe{meas : (E -> Prop) -> Rbar} together
with the fundamental properties making this map a measure.
\begin{lstlisting}
Record measure := mk_measure {
  meas :> (E -> Prop) -> Rbar;
  meas_False : meas (fun _ => False) = 0;
  meas_ge_0 : \forall A, Rbar_le 0 (meas A);
  meas_sigma_additivity : \forall A : nat -> E -> Prop,
    (\forallp n, measurable genE (A n)) -> (\forallp n m x, A n x -> A m x -> n = m) ->
    meas (fun x => \exists n, A n x) = Sup_seq (fun n => sum_Rbar n (fun m => meas (A m)))}.
\end{lstlisting}
The measure is defined as a record.  For the sake of brevity, we want to use it
directly as a function, so we have a coercion (hence the symbol \coqe{:>})
between the type \coqe{measure} and \coqe{(E->Prop)->Rbar}.

The first two properties \coqe{meas_False} and \coqe{meas_ge_0} are
self-explanatory.  Using standard mathematical notations ($\uplus$~denotes the
disjoint union), the $\sigma$-additivity of a map~$\mu$ means that for any
sequence $(A_n)_{n\in\N}$ of pairwise disjoint measurable subsets of~$E$, we
have $\mu\left(\biguplus_{n\in\N}A_n\right)\myequal\sum_{n\in\N}\mu(A_n)$.
Note that infinite summations in~$\Rbarplus$ are formalized as the supremum of
partial sums (see Section~\ref{sec:limit}).

From these fundamental axioms, we prove several other properties of measures
among which {\monotony} ({\ie} $A\subset B\Implies\mu(A)\leq\mu(B)$, for
measurable subsets~$A$ and~$B$), and the weakening of $\sigma$-additivity into
(finite) additivity, for finite unions of pairwise disjoint subsets.  For
instance, the special case of the union of two disjoint subsets simplifies into
\begin{lstlisting}
Lemma measure_additivity :
  \forall (\mu : measure) A B, measurable genE A -> measurable genE B ->
    (\forallp x, A x -> B x -> False) -> \musp (fun x => A x \/ B x) = Rbar_plus (\mu A) (\mu B).
\end{lstlisting}

Another interesting result is the following decomposition of the measure of a
measurable subset \coqe{A : E -> Prop} using a countable partition
\coqe{B : nat -> (E -> Prop)} of the set~\coqe{E}.
\begin{lstlisting}
Lemma measure_decomp :
  \forall (\mu : measure) A (B : nat -> E -> Prop),
    measurable genE A -> (\forallp n, measurable genE (B n)) ->
    (\forallp x, \exists n, B n x) -> (\forallp n p x, B n x -> B p x -> n = p) ->
    \mu A = Sup_seq (fun N => sum_Rbar N (fun n => \musp (fun x => A x /\ B n x))).
\end{lstlisting}
The proof derives directly from $\sigma$-additivity.  A weakened version for
finite partitions is useful to establish additivity of the integral of
{\nonnegative} simple functions in Section~\ref{sec:lint_simple_pos_lin}.

\subsection{Boole's inequality and continuity from below}
\label{sec:measures:boole}

The $\sigma$-additivity and additivity properties described in
Section~\ref{sec:measures:spec_meas} deal with the union of pairwise disjoint
measurable subsets.  When the union is not disjoint, the equality becomes an
inequality, and the resulting {\subadditivity} property is called Boole's
inequality.  The proof path we have followed first addresses the finite case,
then establishes an important intermediate result known as continuity from
below, and finally deals with the infinite case of $\sigma$-{\subadditivity}.

\myskip

Let us first consider finite {\subadditivity}.  It states that for any finite
sequence $(A_n)_{n\in[0..N]}$ of measurable subsets of~$E$, we have
\[
  \mu \left( \bigcup_{n \in [0..N]} A_n \right)
  \leq \sum_{n \in [0..N]} \mu (A_n).
\]
The proof is performed by induction on the parameter~$N$ and uses several
previously proved results, such as additivity and {\monotony} of measures, and
compatibility of measurability with finite union and intersection.  A
specialization for the case~$N=2$, called \coqe{measure_union}, will be handy
in the sequel.

\myskip

The next step is technical, it allows to transform any countable union of
subsets into a pairwise disjoint union, while keeping equal the partial unions.
When the input sequence $(A_n)_{n\in\N}$ is {\nondecreasing}, the new sequence
of pairwise disjoint subsets somehow corresponds to ``nested onion peels'':
$B_0\eqdef A_0$, and for all $n\in\N$, $B_{n+1}\eqdef A_{n+1}\setminus A_n$.
The {\Coq} formalization is {\quasiliteral}.
\begin{lstlisting}
Definition layers : (nat -> E -> Prop) -> nat -> E -> Prop :=
   fun A n => match n with
     | O => A O
     | S n => fun x => A (S n) x /\ ~ A n x
     end.
\end{lstlisting}
We prove several properties of layers, such as compatibility with partial and
countable union ({\ie} $\biguplus_{n\in I}B_n=\bigcup_{n\in I}A_n$ with
\coqe{B := layers A}, for $I=[0..N]$ and $I=\N$), and compatibility with
measurability ({\ie} the layers of a sequence of measurable subsets are
measurable).

\myskip

Our main application of layers and their properties is the continuity from
below of measures.  This results states that for any {\nondecreasing} sequence
$(A_n)_{n\in\N}$ of measurable subsets of~$E$ ({\ie} $A_n\subset A_{n+1}$), we
have
\[
  \mu \left( \bigcup_{n \in \N} A_n \right)
  \leq \lim_{n \to \infty} \mu (A_n).
\]
Note that {\monotony} of measures allows to replace the limit by a supremum
(see Section~\ref{sec:limit}).  Again, the {\Coq} formalization is
straightforward.
\begin{lstlisting}
Definition continuous_from_below : ((E -> Prop) -> Rbar) -> Prop :=
  fun \mu => \forall A : nat -> E -> Prop,
    (\forallp n, measurable genE (A n)) -> (\forallp n x, A n x -> A (S n) x) ->
    \musp (fun x => \exists n, A n x) = Sup_seq (fun n => \musp (A n)).

Lemma measure_continuous_from_below : \forall (\mu : measure), continuous_from_below \mu.
\end{lstlisting}
The proof simply stems from finite additivity and $\sigma$-additivity of
measures, and from careful use of the properties of layers.

\myskip

Finally, let us consider $\sigma$-{\subadditivity}, {\ie} Boole's inequality.
It states that for any sequence $(A_n)_{n\in\N}$ of measurable subsets of~$E$,
we have
\[
  \mu \left( \bigcup_{n \in \N} A_n \right)
  \leq \sum_{n \in \N} \mu (A_n).
\]
It is formalized using \coqe{Sup_seq}.
\begin{lstlisting}
Lemma measure_Boole_ineq :
  \forall (\mu : measure) (A : nat -> E -> Prop), (\forallp n, measurable genE (A n)) ->
    Rbar_le (\mu (fun x => \exists n, A n x)) (Sup_seq (fun n => sum_Rbar n (fun m => \musp (A m)))).
\end{lstlisting}
The proof is an application of continuity from below to the sequence of partial
unions ($B_N\myeqdef\bigcup_{n\in[0..N]}A_n)$.  In {\Coq}, partial unions are
defined using existential quantification that makes the proof process
convenient and fluid.
\begin{lstlisting}
Definition partial_union : (nat -> E -> Prop) -> nat -> E -> Prop :=
  fun A n x => \exists m, (m <== n)%nat /\ A m x.
\end{lstlisting}
Then, the proof resumes by applying finite {\subadditivity} to the
{\nondecreasing} sequence \coqe{partial_}\linebreak[0]%
\coqe{union A}, and using properties of the supremum.

\subsection{Negligible subsets}
\label{sec:measures:negl}

The concepts of \emph{negligible} subset and property satisfied
\emph{almost everywhere} are of major importance in Lebesgue integration
theory.  They are the key ingredients to obtain the positive definiteness
property ({\ie} $\|u\|=0\Implies u=0$) of the norm in~$L^p$ Lebesgue spaces,
which will be the subject of future developments.

A subset~$A$ of~$E$ is said to be negligible for the measure~$\mu$, or simply
$\mu$-negligible, when it is included in a measurable subset~$B$ of measure~0.
\begin{lstlisting}
Definition negligible : (E -> Prop) -> Prop :=
  fun A => \exists B, (\forallp x, A x -> B x) /\ measurable genE B /\ \musp B = 0.
\end{lstlisting}
We prove several simple results about negligible subsets.  For instance,
measurable subsets of measure~0 are negligible, and subsets of negligible
subsets are negligible.  The negligibility of the countable union of negligible
subsets is a bit more challenging; it is a consequence of Boole's inequality.
\begin{lstlisting}
Lemma negligible_union_countable :
  \forall (A : nat -> E -> Prop), (\forallp n, negligible (A n)) -> negligible (fun x => \exists n, A n x).
\end{lstlisting}
This lemma is the only one where we rely on the choice property, see
Section~\ref{sec:basiccoq}.  The reason is as follows.  Given a natural
number~$n$, as we have \coqe{negligible (A n)}, we deduce the existence of
a~$B$ containing~$A_n$ that is both measurable and of measure~0.  But the use
of Boole's inequality and of the measurability of a countable union of sets
require a \emph{sequence} of these~$B$.  So we rely on \coqe{choice} to go from
``for each~$n$, we have a~$B$'' to a sequence of type \coqe{nat -> E -> Prop}
with the expected~properties.

\myskip

A property is said to hold $\mu$-almost everywhere (\coqe{ae}) when its
complement is $\mu$-negligible.
\begin{lstlisting}
Definition ae : (E -> Prop) -> Prop := fun A => negligible (fun x => ~ A x).
\end{lstlisting}
We prove some simple results about $\mu$-almost everywhere properties.  For
instance the countable intersection of properties holding $\mu$-almost
everywhere holds $\mu$-almost everywhere.  This derives from
\coqe{negligible_union_countable}.  An important instantiation of \coqe{ae} is
the equality $\mu$-almost everywhere, used in Section~\ref{sec:lint_p_pos_lin}.
\begin{lstlisting}
Definition ae_eq : (E -> F) -> (E -> F) -> Prop := fun f g => ae (fun x => f x = g x).
\end{lstlisting}

\section{Simple functions}
\label{sec:simple_fun}

Simple functions are real-valued functions that attain only a finite number of
values.  But, unlike step functions used for Riemann sums, each value may be
taken here on a {\nonconnected} subset.

This is a very simple mathematical definition, but it will require some proof
engineering to have a usable formal definition.  Another mathematical
equivalent definition is that a simple function is a finite linear combination
of indicator functions, and can be expressed as
\begin{equation}
  \label{eq:def:simple-fun}
  f = \sum_{y \in f (E)} y \times \charac{\preimage{f}{y}},
\end{equation}
where~$\charac{}$ is the characteristic function (see
Section~\ref{sec:charac}).  This definition is impractical in {\Coq} as it sums
over~$f(E)$ that may be infinite in general.  Only the property of~$f$ makes
this subset finite.  We choose to have a data structure that allows us to
access the possible values, in order to be able to compute the integral of
simple functions, and we choose to have them as a list.  Indeed, the {\Coq}
{\List} library is rather comprehensive, even if not perfectly suited for our
use.  We also finally choose to have simple functions of type \coqe{E -> R} and
not \coqe{E -> Rbar}; this is discussed in Section~\ref{sec:discuss:simplefun}.

We consider an ambient set~\coqe{E} now required to be inhabited.  The empty
case is not of interest here, and it would mean empty lists that make the
following functions fail.  Instead of having additional hypotheses on the
lists, it was easier not to consider empty types.  Given a function and a list,
the property \coqe{finite_vals} states that the values taken by the function
belong to the list.
\begin{lstlisting}
Definition finite_vals : (E -> R) -> list R -> Prop := fun f l => \forall x, In (f x) l.
\end{lstlisting}
Note that this list is far from unique: the elements may be in any order, can
be taken several times, and useless values may be in the list.  Hence, the need
for a canonical list that is computed in Section~\ref{sec:canonizer}, in order
to integrate {\nonnegative} simple functions, as described in
Section~\ref{sec:lint_simple}.  The~positive linearity of the integral is shown
in Section~\ref{sec:lint_simple_pos_lin}.

\subsection{Canonical representation}
\label{sec:canonizer}

As explained above, the property \coqe{finite_vals} does not specify a unique
list.  To enforce uniqueness, we need a strictly sorted list, with only the
useful values.
\begin{lstlisting}
Definition finite_vals_canonic : (E -> R) -> list R -> Prop :=
  fun f l => (LocallySorted Rlt l) /\ (\forallp y, In y l -> \exists x, f x = y) /\ (\forallp x, In (f x) l).
\end{lstlisting}
where \coqe{LocallySorted P} (from the {\Coq} standard library) is the
inductive definition of a sorted list using the~\coqe{P} relation.  Here, we
require the strict order \coqe{Rlt} to prevent duplicates.

The related proofs are then threefold.  First, we need to prove that only one
list fits the requirement.
\begin{lstlisting}
Lemma finite_vals_canonic_unique :
  \forall f l1 l2, finite_vals_canonic f l1 -> finite_vals_canonic f l2 -> l1 = l2.
\end{lstlisting}
The proof is not difficult, but slightly tedious.  An intermediate lemma states
that if two lists have the same elements (using \coqe{In}) and are both
\coqe{LocallySorted} with \coqe{Rlt}, then they are equal.

Second, to recover the fact that our simple functions are indeed a finite
linear combination of indicator functions, we also prove that
\coqe{finite_vals_canonic $f$ $\ell$} implies the same equality
as~\eqref{eq:def:simple-fun}, but for~$y$ in the list:
$f=\sum_{y\in\ell}y\times\charac{\preimage{f}{y}}$.
\begin{lstlisting}
Lemma finite_vals_sum_eq :
  \forall (f : E -> R) l, finite_vals_canonic f l ->
    \forall x, f x = sum_Rbar_map l (fun y => y * (charac (fun z => f z = y) x)).
\end{lstlisting}

\myskip

Last but not least, we need to be able to build this canonical list using
several intermediate~steps.
\begin{lstlisting}
Fixpoint select {E : Type} (P : E -> Prop) (l : list E) : list E :=
  match l with
  | nil => nil
  | y :: l1 => match (excluded_middle_informative (P y)) with
    | left _ => y :: select P l1
    | right _ => select P l1
    end
  end.

Definition RemoveUseless : \forall {E F : Type}, (E -> F) -> list F -> list F :=
  fun E F f l => select (fun y => \exists x, f x = y) l.

Definition canonizer : (E -> R) -> list R -> list R :=
  fun f l => sort Rle (RemoveUseless (nodup Req_EM_T l) f).
\end{lstlisting}

Let us explain these functions.  The \coqe{select} function takes advantage of
the \coqe{excluded_middle_}\linebreak[0]%
\coqe{informative} axiom to select the values of a list having a given
property.  The \coqe{RemoveUseless} function then allows us to select only the
useful values of the list (such that there exists a preimage to it).  The
\coqe{nodup} function from the {\Coq} standard library removes duplicates (the
decidability of equality on real numbers is given).  We redefined the sort
function and call it with the {\nonstrict} order \coqe{Rle}.

The canonizer function is then a successive call to \coqe{nodup},
\coqe{RemoveUseless} and \coqe{sort}.  Note that these operations are actually
commuting, thus any ordering would have been correct.  We choose the one that
eases the proofs.  In particular, \coqe{sort} is the last called function as it
will imply an easy proof that the final list is sorted.  The function
\coqe{nodup} is called first as few lemmas are available on~it.

The correctness of this canonizer is then proved.
\begin{lstlisting}
Lemma finite_vals_canonizer :
  \forall f l, finite_vals f l -> finite_vals_canonic f (canonizer f l).
\end{lstlisting}

For instance, consider the case of the characteristic function of some proper
subset~$A$ (distinct from the empty set and from the full set, thus assuming
that type~\coqe{E} is neither empty, nor representing a singleton).  This
function actually takes the values~0 and~1 (see Section~\ref{sec:charac}): it
is a simple function.  Starting with any list of real numbers containing~0
and~1, {\eg} \coqe{l := [1; 0; 0; 2]}, we may show the property
\coqe{finite_vals (charac A) l}.  And thus, from the previous lemma, we have
\coqe{finite_vals_canonic (charac A) (canonizer (charac A) l)}, where the
canonized list is simply \coqe{[0; 1]}.

\subsection{Integration of {\nonnegative} simple functions}
\label{sec:lint_simple}

Following the definition of simple functions, we retain those for which
preimages of singletons are measurable, and thus admit a measure, possibly
infinite.  Those measurable simple functions are collected into the
set~$\calSF$, and the subset of {\nonnegative} ones is denoted~$\calSFplus$.
The needed tools for integrating in~$\calSFplus$ are sums on~$\Rbar$ as defined
in Section~\ref{sec:sum_Rbar}, and a measure~$\mu$ as defined in
Section~\ref{sec:measures}.  The integral of $f\in\calSFplus$ is defined by
\begin{equation}
  \label{eq:def:lint_simple}
  \intSFplus f \, d\mu
  \eqdef \sum_{y \in f (E)} y \times \mu \left( \preimage{f}{y} \right).
\end{equation}

We first need to specify simple functions of~$\calSF$, that have measurable
preimages.
\begin{lstlisting}
Definition SF_aux : ((E -> Prop) -> Prop) -> (E -> R) -> list R -> Prop :=
  fun genE f lf => finite_vals_canonic f lf /\ (\forallp y, measurable genE (fun x => f x = y)).

Definition SF : ((E -> Prop) -> Prop) -> (E -> R) -> Set :=
  fun genE f => {lf | SF_aux genE f lf}.
\end{lstlisting}
Note that the list is in \coqe{Set} as we need to get a hand on it to compute
the integral.  A weak existential is not strong enough for our purpose.  In
{\Coq}, the notation \coqe!{x | P x}! means \emph{there exists \coqe{x} such
  that \coqe{P x}}, but the type is \coqe{Set}.  This means it is a strong
existential, {\ie} it is possible to pick up the witness~\coqe{x}.

Note also that since singletons are Borel subsets of~$\R$ (as closed subsets),
we are able to prove measurability of functions in~$\calSF$.
\begin{lstlisting}
Lemma SF_aux_measurable_fun: \forall genE f l, SF_aux genE f l -> measurable_fun_R f.
\end{lstlisting}

Then, the definition of the integral in~$\calSFplus$ is straightforward from a
proof of type~\coqe{SF}.
\begin{lstlisting}
Definition af1 : (E -> R) -> Rbar -> Rbar :=
  fun f y => Rbar_mult y (\mu (fun x => Finite (f x) = y))).

Definition LInt_SFp : \forall genE, \forall (f : E -> R), SF genE f -> Rbar :=
  fun f Hf => let lf := proj1_sig Hf in sum_Rbar_map lf (af1 f).
\end{lstlisting}
Note the required hypothesis~\coqe{Hf} that encompasses both the proof that~$f$
is a valid simple function, and the list witness~$\ell$ on which the definition
depends.  Then, \coqe{proj1_sig} returns the first part of this proof, that is
the list \coqe{lf}, in order to sum on it.  This dependent type is only inside
the library and is not to be used outside: final users will make use only of a
total function for the Lebesgue~integral.  This limited use of dependent types
has not proved inconvenient.

We first prove that the value of the integral does not depend on the chosen
list/proof (\coqe{Hf1} and \coqe{Hf2} in the next lemma).
\begin{lstlisting}
Lemma LInt_SFp_correct :
  \forall f (Hf1 Hf2 : SF genE f), non_neg f -> LInt_SFp f Hf1 = LInt_SFp f Hf2.
\end{lstlisting}

A first easy example of integration is the relationship between this integral
and the characteristic function.  The characteristic function has two possible
values (0 and 1), so it is a simple~function.  When the subset~$A$ is
measurable, $\charac{A}$~belongs to~$\calSFplus$, and its integral is, as
expected, the measure of~$A$.
\begin{lstlisting}
Lemma LInt_SFp_charac :
  \forall A (HA : measurable genE A), LInt_SFp (charac A) (SF_charac A HA) = \musp A.
\end{lstlisting}

\subsection{Linearity of the integral of simple functions}
\label{sec:lint_simple_pos_lin}

Then comes a proof that is unexpectedly complex, the additivity of the integral
in~$\calSFplus$%
\mydisplaymaths{\intSFplus(f+g)\,d\mu=\intSFplus f\,d\mu+\intSFplus g\,d\mu}

Alternate proofs are available, {\eg} see~\cite{rud:rca:87,gh:mip:13}, but were
not considered in this work.  We~choose a proof using a change of variable,
from the sum of values taken by each function~$f$ and~$g$ to the values taken
by their sum~$f+g$.  The main difficulty is that the canonical
list~$\ell_{f+g}$ associated with~$f+g$ has nothing to do with any kind of
``addition'' of the lists~$\ell_f$ and~$\ell_g$ associated with~$f$ and~$g$.

\myskip

The first stage is a lemma stating that~$\calSF$ is closed under addition.
\begin{lstlisting}
Lemma SF_plus : \forall f (Hf : SF genE f) g (Hg : SF genE g), SF genE (fun x => f x + g x).
\end{lstlisting}
For that, we rely on
\begin{lstlisting}
Definition cartesian_Rplus : list R -> list R -> list R :=
  fun l1 l2 => flat_map (fun a1 => (map (fun a2 => a1 + a2) l2)) l1.
\end{lstlisting}
that gathers all possible sums from two lists.  When applied to~$\ell_f$
and~$\ell_g$, the result may be too large a list, but no useful value is
missing.  Thus, we may strip unwanted values by applying the previously defined
canonizer(see Section~\ref{sec:canonizer}).

\myskip

The second stage is a couple of lemmas coming from the fact that the subsets
$\preimage{f}{y}$ for $y\in f(E)$ constitute a partition of the domain~$E$ of
the function~$f$.  First, a specialization of the finite version of the lemma
\coqe{measure_decomp} (see Section~\ref{sec:measures:spec_meas}) for preimages
by functions~$f$ and~$g$ provides
\begin{lstlisting}
Lemma SFp_decomp :
  \forall f g lf lg y, SF_aux genE f lf -> SF_aux genE g lg ->
    \mu (fun x => f x = y) = sum_Rbar_map lg (fun z => \musp (fun x => f x = y /\ g x = z)).
\end{lstlisting}
Note that this result is first proved with the assumption that~$y$ is actually
a value taken by~$f$.  But~this premise can be dropped as for all other values
of~$y$, the equality to show simplifies into the trivial $0=0$.

Then, the result is lifted to the integral level,
\[
  \sum_{y \in f (E)} y \times \mu \left( \preimage{f}{y} \right)
  = \sum_{y \in f (E)} y \left( \sum_{z \in g (E)}
    \mu \left( \preimage{f}{y} \cap \preimage{g}{z} \right) \right),
\]
which is formalized as
\begin{lstlisting}
Lemma LInt_SFp_decomp :
  \forall f g lf lg, SF_aux genE f lf -> SF_aux genE g lg ->
    (* LInt_SFp f H = *) sum_Rbar_map lf (af1 f) = sum_Rbar_map lf
      (fun y => Rbar_mult y (sum_Rbar_map lg (fun z => \musp (fun x => f x = y /\ g x = z)))).
\end{lstlisting}

\myskip

The third stage consists in applying the latter lemma to justify the rewritings
in the equations~below.  First, for both integrals in the sum.
\begin{eqnarray*}
  \intSFplus f \, d\mu + \intSFplus g \, d\mu
  & = & \sum_{y \in f (E)} y \, \mu \left( \preimage{f}{y} \right)
        + \sum_{z \in g (E)} z \, \mu \left( \preimage{g}{z} \right)\\
  & = & \sum_{y \in f (E)} \sum_{z \in g (E)} y \,
        \mu \left( \preimage{f}{y} \cap \preimage{g}{z} \right) \\
  && + \sum_{z \in g (E)} \sum_{y \in f (E)} z \,
     \mu \left( \preimage{g}{z} \cap \preimage{f}{y} \right)\\
  & = & \sum_{y \in f (E)} \sum_{z \in g (E)} (y + z) \,
        \mu \left( \preimage{f}{y} \cap \preimage{g}{z} \right).
\end{eqnarray*}
And then, for the integral of the sum, where the lemma is applied with
\coqe{f}~$=f+g$ and \coqe{g}~$=f$.
\begin{eqnarray*}
  \intSFplus (f + g) \, d\mu
  & = & \sum_{t \in (f + g) (E)} t \, \mu \left( \preimage{(f + g)}{t} \right)\\
  & = & \sum_{t \in (f + g) (E)} t \, \sum_{y \in f (E)}
        \mu \left( \preimage{(f + g)}{t} \cap \preimage{f}{y} \right)\\
  & = & \sum_{y \in f (E)} \sum_{t \in (f + g) (E)} t \,
        \mu \left( \preimage{f}{y} \cap \preimage{(f + g)}{t} \right).
\end{eqnarray*}

\myskip

Finally, the last step of the additivity proof is to connect both sets of
equalities by establishing the following ``horrible'' change of variable
formula
\begin{multline*}
  \sum_{z \in g (E)} (y + z) \,
  \mu \left( \preimage{f}{y} \cap \preimage{g}{z} \right)\mybreakmaths
  = \sum_{t \in (f + g) (E)} t \,
  \mu \left( \preimage{f}{y} \cap \preimage{(f + g)}{t} \right),
\end{multline*}
that is formalized as
\begin{lstlisting}
Lemma sum_Rbar_map_change_of_variable :
  \forall f g lf lg y, SF_aux genE f lf -> SF_aux genE g lg ->
    let lfpg := canonizer (fun x => f x + g x) (cartesian_Rplus lf lg) in
    sum_Rbar_map lg (fun z => Rbar_mult (y + z) (\mu (fun x => f x = y /\ g x = z))) =
      sum_Rbar_map lfpg
        (fun t (* = y + z *) => Rbar_mult t (\mu (fun x => f x = y /\ f x + g x = t))).
\end{lstlisting}
The key ingredient here is that sums may be restricted to only their {\nonzero}
terms, which makes the change of variable $z\mapsto t=y+z$ (for fixed~$y$) a
bijection.

An interesting point is that this lemma is hardly explicit in mathematical
textbooks and we had to puzzle it out to fulfill the proof.  We had to write it
explicitly as it was a key point in our design choice for simple functions, see
Section~\ref{sec:discuss:simplefun}.

Ultimately, we end up with similar double summations, and we are able to prove
the additivity of the integral in~$\calSFplus$.
\begin{lstlisting}
Lemma LInt_SFp_plus :
   \forall f (Hf : SF genE f) g (Hg : SF genE g), non_neg f -> non_neg g ->
    let Hfpg := SF_plus f Hf g Hg in
    LInt_SFp (fun x => f x + g x) Hfpg = Rbar_plus (LInt_SFp f Hf) (LInt_SFp g Hg).
 \end{lstlisting}

\myskip

As a break, we establish the compatibility of the integral in~$\calSFplus$ with
{\nonnegative} scaling.
\begin{lstlisting}
Lemma LInt_SFp_scal :
  \forall f (Hf : SF genE f) a, non_neg f -> 0 <== a ->
    let Haf := SF_scal f Hf in
    LInt_SFp (fun x => a * f x) Haf = Rbar_mult a (LInt_SFp f Hf).
\end{lstlisting}
This calls for a proof~\coqe{Haf} that a simple function multiplied by a scalar
is indeed a simple function.  Then, we only need to require that both the
scalar and the function are {\nonnegative} to ensure that
$\intSFplus a\times f\,d\mu=a\times\intSFplus f\,d\mu$.  Then, {\monotony} of
the integral in~$\calSFplus$ is a direct consequence of additivity, since the
relation $g=f+(g-f)$ holds in~$\calSFplus$ when $f\leq g$.

\section{Integration of {\nonnegative} functions}
\label{sec:lint_p}

Let us now consider functions of type \coqe{E -> Rbar}, that may take an
infinite number of (possibly infinite) values.  The set of measurable functions
to~$\Rbar$ is denoted by~$\calM$, and its subset of {\nonnegative} functions
by~$\calMplus$.  The key idea for the definition of the integral in~$\calMplus$
is to use approximations from below by simple functions in~$\calSFplus$, and
surprisingly, we benefit from the use of computer arithmetic.

The integral is presented in Section~\ref{sec:lint_p_def} together with some
preliminary properties.  Then, Section~\ref{sec:bigth:bele_tt} is devoted to
the crucial {\BeppoLeviMonotConvTh}.  Adapted sequences are defined in
Section~\ref{sec:lint_p_adap_seq}.  Linearity and other properties of the
integral are displayed in Section~\ref{sec:lint_p_pos_lin}.  Finally,
Section~\ref{sec:bigth:fale_th} is devoted to {\FatouLem}, the other major
result on the integral for {\nonnegative} functions.

\subsection{Definition and first properties}
\label{sec:lint_p_def}

We now want to define the Lebesgue integral for {\nonnegative} integrable
functions.  The mathematical definition is
\[
  \forall f \in \calMplus,\quad
  \intMplus f \, d\mu
  \EQDEF
  \sup_{\substack{\psi \in \calSFplus\\\psi \leq f}} \intSFplus \psi \, d\mu
\]
where the supremum is taken for {\nonnegative} measurable simple
functions~$\psi$ less than or equal to~$f$ pointwise, and where the integral
in~$\calSFplus$ is defined in Section~\ref{sec:lint_simple}.

Keeping on with total functions, we prescribe a value whatever the input
function~$f$, with a {\Coq} definition quite similar to the mathematical one.
\begin{lstlisting}
Definition LInt_p : (E -> Rbar) -> Rbar :=
  fun f => Rbar_lub (fun z : Rbar => \exists (\psi : E -> R) (H\psi : SF genE \psi),
    non_neg \psisp /\ (\forallp x, Rbar_le (\psi x) (f x)) /\ LInt_SFp \musp \psisp H\psi = z).
\end{lstlisting}
The supremum is taken on a subset of extended reals~$z$ such that there exists
a simple function \coqe{\psi} less than or equal to~$f$ having~$z$ for
integral.

\myskip

The first thing to prove is that this newly-defined integral is the same as the
already-defined integral when the function is a simple function, {\ie}
$\intMplus f\,d\mu$ is equal to $\intSFplus f\,d\mu$ for all~$f$
in~$\calSFplus$, and in {\Coq}
\begin{lstlisting}
Lemma LInt_p_SFp_eq :
  \forall (f : E -> R) (Hf : SF genE f), non_neg f -> LInt_p f = LInt_SFp \musp f Hf.
\end{lstlisting}

Then comes the {\monotony} of the integral.
\[
  \forall f, g \in \calMplus,\quad
  f \leq g \IMPLIES \intMplus f \, d\mu \LEQ \intMplus g \, d\mu.
\]
The {\Coq} translation becomes
\begin{lstlisting}
Lemma LInt_p_monotone :
  \forall (f g : E -> Rbar), (\forallp x, Rbar_le (f x) (g x)) -> Rbar_le (LInt_p f) (LInt_p g).
\end{lstlisting}
Indeed, the least upper bound (LUB) in the definition of the total function is
enough to ensure {\monotony} for any functions~$f$ and~$g$, not only for the
{\nonnegative} and measurable ones as in the mathematical statement.

Another easy result is about the multiplication by a {\nonnegative} scalar
value.
\begin{lstlisting}
Lemma LInt_p_scal_finite :
  \forall (f : E -> Rbar) (a : R), 0 <== a ->
    LInt_p (fun x => Rbar_mult a (f x)) = Rbar_mult a (LInt_p f).
\end{lstlisting}
As before, there is no assumption on the fact that~$f$ is {\nonnegative}.

The following extensionality result instantiated for restricted functions has
proved useful.  It~states that when functions are equal on a measurable subset,
then the integral of their restriction to that subset are equal.  This is
hardly mentioned in mathematics.  As before, there is no requirement on the
properties of~$A$, $f$ and~$g$.  The total function \coqe{LInt_p} gives
something that is the same in both cases, even for {\nonmeasurable} functions.
\begin{lstlisting}
Lemma LInt_p_when_charac :
  \forall f g (A : E -> Prop), (\forallp x, A x -> f x = g x) ->
    LInt_p (fun x => Rbar_mult (f x) (charac A x)) =
      LInt_p (fun x => Rbar_mult (g x) (charac A x)).
\end{lstlisting}

\subsection{The {\BeppoLeviTh}}
\label{sec:bigth:bele_tt}

The~{\BeppoLeviTh} (see Textbook Theorem~\ref{th:beppo-levi},
page~\pageref{th:beppo-levi}), also known as the monotone convergence theorem,
is one of the most fundamental results in measure and integration theories.  It
states that for any sequence $(f_{n})_{n \in \N}$ of pointwise {\nondecreasing}
and {\nonnegative} measurable functions ({\ie} in~$\calMplus$), the pointwise
limit $\lim_{n \to \infty} f_n$ (which actually equals $\sup_{n \in \N} f_n$,
see Section~\ref{sec:limit}) is also in~$\calMplus$.  This property is proved
using standard properties of measurable functions such as {\monotony}, and is
not particularly challenging.  The {\BeppoLeviTh} also states the following
integral-limit exchange formula
\[
  \intMplus \sup_{n \in \N} f_n \, d\mu
  = \sup_{n \in \N} \intMplus f_n \, d\mu
\]
which is stated in {\Coq} as
\begin{lstlisting}
Lemma Beppo_Levi :
  \forall (f : nat -> E -> Rbar), (\forallp x n, Rbar_le (f n x) (f (S n) x)) ->
    (\forallp n, non_neg (f n)) -> (\forallp n, measurable_fun_Rbar genE (f n)) ->
    LInt_p \musp (fun x => Sup_seq (fun n => f n x)) = Sup_seq (fun n => LInt_p \musp (f n)).
\end{lstlisting}

The proof of this equality is technical, and relies on a wide variety of
previously proved results.  It can be divided into two inequalities.  The easy
one, $\sup_{n\in\N}\intMplus f_n\,d\mu\leq\intMplus\sup_{n\in\N}f_n\,d\mu$, is
proved using {\monotony} of the integral, and fundamental properties of the
supremum.  The other one,
\begin{equation}
  \label{eq:BL-hard-ineq}
  \intMplus \sup_{n \in \N} f_n \, d\mu
  \leq \sup_{n \in \N} \intMplus f_n \, d\mu,
\end{equation}
is more intricate.  The first step of the proof is to show that for any
$\psi\in\calSFplus$ less than or equal to $\sup_{n\in\N}f_n$, and any number
$0<a<1$, we have the bound
\begin{equation}
  \label{eq:BL-majoration}
  \intMplus \psi \, d\mu
  \leq \frac{1}{a} \sup_{n \in \N} \intMplus f_n \, d\mu.
\end{equation}

For that purpose, the subsets $A_n=\{x\in E\,|\,a\psi\leq f_n\}$ are first
shown to be {\nondecreasing} and measurable, the latter coming from the
measurability of functions~$a\psi-f_n$.  Then, they are shown to cover the full
set~$E$, which is stated in {\Coq} as \coqe{\forall x, \exists n, A n x}, and
the existential is exhibited as a rank~$N$ above which we have
$a\psi(x)\leq f_n(x)$.  Then, the proof of Equation~\eqref{eq:BL-majoration}
relies on continuity from below of the measure (see
Section~\ref{sec:measures:boole}), measurability of simple functions (see
Section~\ref{sec:lint_simple}), linearity properties of the integral
in~$\calSFplus$ (see Section~\ref{sec:lint_simple_pos_lin}), and {\monotony}
and compatibility with characteristic functions for the integral in~$\calMplus$
(see Section~\ref{sec:lint_p_def}).

Finally, the inequality~\eqref{eq:BL-hard-ineq} is obtained by taking
in~\eqref{eq:BL-majoration} the limit as~$a$ goes up to~1, and from the
definition of the integral in~$\calMplus$ (see Section~\ref{sec:lint_p_def})
and properties of the supremum.

This concludes the proof of the~{\BeppoLeviTh}.

\subsection{Adapted sequences}
\label{sec:lint_p_adap_seq}

As for simple functions, a real difficulty is the additivity property of the
integral in~$\calMplus$%
\mydisplaymaths{\intMplus(f+g)\,d\mu=\intMplus f\,d\mu+\intMplus g\,d\mu} Given
the definition of the Lebesgue integral (see Section~\ref{sec:lint_p_def}), the
common proof relies on adapted sequences in~$\calSFplus$.

An adapted sequence for a function~$f$ is a pointwise {\nondecreasing} sequence
of {\nonnegative} functions that is pointwise converging from below toward~$f$.
\begin{lstlisting}
Definition is_adapted_seq : (E -> Rbar) -> (nat -> E -> R) -> Prop :=
  fun f \psisp => (\forallp n, non_neg (\psi n)) /\ (\forallp x n, \psi n x <== \psisp (S n) x) /\
    (\forallp x, is_sup_seq (fun n => \psisp n x) (f x)).
\end{lstlisting}
In our case, the adapted sequences of interest are the measurable simple
functions of~$\calSF$.  We~then deduce that the sequence of integrals of such a
sequence converges toward the integral of~$f$ from~below.
\begin{lstlisting}
Definition SF_seq : (nat -> E -> R) -> Set := fun \psisp => \forall n, SF genE (\psi n).

Lemma LInt_p_with_adapted_seq :
  \forall f \psi, is_adapted_seq f \psisp -> SF_seq genE \psisp ->
    Sup_seq (fun n => LInt_p \musp (\psi n)) = LInt_p \musp f.
\end{lstlisting}

\myskip

Having this definition and its link to the integral is far from enough.  We
need to have an adapted sequence corresponding to any function in~$\calMplus$.
This building is quite easy in mathematics: for~$f$ {\nonnegative}, the chosen
adapted sequence is
\begin{equation}
  \label{eq:adap-seq}
  \fhi_n (x) \eqdef \left\{
    \begin{array}{ll}
      \displaystyle
      \frac{\floor{2^n f (x)}}{2^n} & \mbox{when } f (x) < n,\\
      n & \mbox{otherwise}.
    \end{array}
  \right.
\end{equation}

We began by translating literally this definition.  Then, we tried to prove
that the sequence is {\nondecreasing}, and so on.  One of the authors then
noticed that such proofs were already done and available in the library
{\Flocq}~\cite{BolMel11} dedicated to computer arithmetic.  {\Flocq} is a
formalization of floating-point arithmetic that provides a comprehensive
library of theorems on a multi-radix multi-precision arithmetic.  It also
supports efficient numerical computations inside {\Coq}.  As it aims at
genericity, even computer arithmetic formats are abstract to encompass
floating-point and fixed-point arithmetics and many proved results also hold in
fixed-point arithmetic.  Seen from a computer science point of view, the
definition of~$\fhi_n$ in Equation~\eqref{eq:adap-seq} indeed relies on a
fixed-point rounding downward with a least significant bit (lsb) of~$-n$.  It
is formalized in {\Coq} as
\begin{lstlisting}
Definition mk_adapted_seq (* = \phinsp *) : nat -> E -> R :=
  fun n x => match (Rbar_le_lt_dec (INR n) (f x)) with
    | left _ => INR n
    | right _ => round radix2 (FIX_exp (-Z.of_nat n)) Zfloor (f x)
    end.
\end{lstlisting}

Many proofs related to inequalities (such as $(\fhi_n)_{n\in\N}$ is bounded and
{\nondecreasing}) are really smooth, relying on the support library of
{\Flocq}.  For instance, the theorem $\fhi_n(x)\le f(x)$ is a split whether
$f(x)<n$, followed by a call to a property of the rounding downward.

A more involved example lies in the proof of $\fhi_n(x)\le\fhi_{n+1}(x)$.  We
first split depending whether $f(x)<n$ or~$n+1$, and handle three simple cases.
The most difficult case should be to prove that
\[
  \frac{\floor{2^n f (x)}}{2^n} \le \frac{\floor{2^{n+1} f (x)}}{2^{n+1}}
\]
when $f(x)<n$.  Indeed, a direct proof does not seem so straightforward.
Taking the computer arithmetic point of view, it becomes
$\bigtriangledown_n(x)\le\bigtriangledown_{n+1}(x)$, with~$\bigtriangledown_n$
the rounding down in fixed-point arithmetic with least significant bit
(lsb)~$n$.  Then, we rely on the following floating-point theorem: if $u\le v$
and $u\in\F$, then $u\le\circ(v)$ for any reasonable format~$\F$ and
rounding~$\circ$.  Applying it, there is left to prove both that
$\bigtriangledown_n(x)\le x$ (trivial by the properties of the rounding down)
and that~$\bigtriangledown_n(x)$ fits in the fixed-point format with least
significant bit~$-n-1$ (easy as it has a lsb of~$-n$).  The proof of this
subcase has been done in 10~lines of standard {\Coq}.

Even the convergence was eased by existing {\Flocq} error lemmas.  The main
result states that the~$\fhi_n$ are indeed an adapted sequence.
\begin{lstlisting}
Lemma mk_adapted_seq_is_adapted_seq : is_adapted_seq f mk_adapted_seq.
\end{lstlisting}

\myskip

The part left to prove is that the~$\fhi_n$ are in~$\calSF$.  The first thing
to prove is that the preimages of~$\fhi_n$ are measurable subsets.
\begin{lstlisting}
Lemma mk_adapted_seq_SF_aux : \forall n y, measurable genE (fun x => mk_adapted_seq n x = y).
\end{lstlisting}
There are several proof paths.  The chosen one is to prove that the subset is
either empty (so measurable), or $y\le n$.  We also prove that~$y$ is a
fixed-point number with lsb~$n$.  Then, we have
\[
  \fhi_n (x) = y \EQUIV y \le f (x) < \succ (y)
\]
where~$\succ$ is the successor in fixed-point arithmetic, meaning the next
number in the format with lsb~$n$.  This subset is measurable as~$f$ is
measurable, and from the properties of Borel subsets of
Section~\ref{sec:gen_equiv}.  There are some special cases related to~$\infty$
and to $y=n$ that are tedious but easy.

Last is to prove that the~$\fhi_n$ only take a finite number of values, making
them valid simple functions.
\begin{lstlisting}
Lemma mk_adapted_seq_SF : SF_seq genE mk_adapted_seq.
\end{lstlisting}
The mathematical definition is clear, but this is the first proof of this kind
(the only previously proved simple function is the characteristic function).
So we have to build by hand the list of all possible values of~$\fhi_n$: first
the list of integers~$i$ with $0\le i\le n$, and then the list of all the
$i/2^n$ for $0\le i\le n2^n$.  This kind of proof could clearly be automated,
or simplified by dedicated lemmas if need be.  From this very generic list, we
only have to apply the canonizer defined in Section~\ref{sec:canonizer} to~$f$.
Then, we end up the proof, relying on the various properties of the canonizer,
the fixed-point rounding, and the measurability above.

\myskip

To conclude, we have defined explicitly an adapted sequence that we may give to
the theorem \coqe{LInt_p_with_adapted_seq}, thus providing an explicit formula
for the integral in~$\calMplus$%
\mydisplaymaths{\intMplus f\,d\mu=\sup_{n\in\N}\intMplus\fhi_n\,d\mu}
\begin{lstlisting}
Lemma LInt_p_with_mk_adapted_seq :
  \forall f, non_neg f -> measurable_fun_Rbar genE f ->
    Sup_seq (fun n => LInt_p \musp (mk_adapted_seq f n)) = LInt_p \musp f.
\end{lstlisting}

\subsection{Linearity and other properties of the integral}
\label{sec:lint_p_pos_lin}

We present now some theorems about the integration of {\nonnegative} measurable
functions that we consider essential for a library user.  They are all
consequences of the {\BeppoLeviMonotConvTh} (see
Section~\ref{sec:bigth:bele_tt}).  They are gathered in
Table~\ref{tab:pos_lin}.

\renewcommand\cellalign{{{p{0.39\textwidth}}}}
\begin{longtable}{|l|l|}
  \hline
  {\Coq} statements & Mathematical statements\\\hline
\begin{lstlisting}
Lemma LInt_p_plus :
  \forall f g,
    non_neg f -> measurable_fun_Rbar genE f ->
    non_neg g -> measurable_fun_Rbar genE g ->
    LInt_p \musp (fun x => Rbar_plus (f x) (g x)) =
      Rbar_plus (LInt_p \musp f) (LInt_p \musp g).
\end{lstlisting}
& \makecell{%
  $\forall f,g\in\calMplus,$\\
  $\intMplus(f+g)\,d\mu\EQ$\\
  \mbox{}\hfill$\intMplus f\,d\mu\PLUS\intMplus g\,d\mu.$}
  \\\hline
\begin{lstlisting}
Lemma LInt_p_scal :
  \forall a f, Rbar_le 0 a ->
    non_neg f -> measurable_fun_Rbar genE f ->
    LInt_p \musp (fun x => Rbar_mult a (f x)) =
      Rbar_mult a (LInt_p \musp f).
\end{lstlisting}
& \makecell{%
  $\forall a\in\Rbarplus,\;\forall f\in\calMplus,$\\
  $\intMplus(a\times f)\,d\mu\EQ a\times\intMplus f\,d\mu.$}
  \\\hline
\begin{lstlisting}
Lemma LInt_p_ae_definite :
  \forall f,
    non_neg f -> measurable_fun_Rbar genE f ->
    (ae_eq \musp f (fun _ => 0) <-> LInt_p \musp f = 0).
\end{lstlisting}
& \makecell{%
  $\forall f\in\calMplus,$\\
  $f\eqae{\mu}0\,\Equiv\,\intMplus f\,d\mu=0.$}
  \\\hline
\begin{lstlisting}
Lemma LInt_p_decomp :
  \forall f A, measurable genE A ->
    non_neg f -> measurable_fun_Rbar genE f ->
    LInt_p \musp f = Rbar_plus
      (LInt_p \musp (fun x => Rbar_mult (f x)
        (charac A x)))
      (LInt_p \musp (fun x => Rbar_mult (f x)
        (charac (fun y => ~ A y) x))).
\end{lstlisting}
& \makecell{%
  $\forall f\in\calMplus,\;\forall A\ \measurable,$\\
  $\intMplus f\,d\mu\EQ\intMplus(f\times\charac{A})\,d\mu$\\
  \mbox{}\hfill$\PLUS\intMplus(f\times\charac{A^c})\,d\mu.$}
  \\\hline
\begin{lstlisting}
Lemma LInt_p_ae_eq_compat:
  \forall f g,
    non_neg f -> measurable_fun_Rbar genE f ->
    non_neg g -> measurable_fun_Rbar genE g ->
    ae_eq \musp f g ->
    LInt_p \musp f = LInt_p \musp g.
\end{lstlisting}
& \makecell{%
  $\forall f,g\in\calMplus,$\\
  $f\eqae{\mu}g\,\Implies$\\
  \mbox{}\hfill$\intMplus f\,d\mu\EQ\intMplus g\,d\mu.$}
  \\\hline
\caption{Linearity and other properties of the integral in~$\calMplus$.}
\label{tab:pos_lin}
\end{longtable}

The first entry (\coqe{LInt_p_plus}) is for additivity of the integral
in~$\calMplus$.  The proof is rather smooth, now that additivity of the
integral in~$\calSFplus$ (see Section~\ref{sec:lint_simple_pos_lin}), the
monotone convergence (see Section~\ref{sec:bigth:bele_tt}), and existence of an
adapted sequence (see Section~\ref{sec:lint_p_adap_seq}) are established.
The~second entry (\coqe{LInt_p_cal}) generalizes the {\nonnegative} scaling
property (see Section~\ref{sec:lint_p_def}) to the infinite case too.  Both
will be significant ingredients of the linearity of the Lebesgue integral for
arbitrary sign functions, to build the structure of vector space of the
integrable functions (out of the scope of this paper).

The remaining entries actually follow from the first two linearity results.
The characterization of zero-integral functions (\coqe{LInt_p_ae_definite})
relies on the scaling property and the compatibility result with characteristic
functions (see Section~\ref{sec:lint_p_def}), while the decomposition on a
partition (\coqe{LInt_p_decomp}) relies on the additivity property.  Finally,
the compatibility result with almost equality (\coqe{LInt_p_ae_}\linebreak[0]%
\coqe{eq_compat}) relies on the last two.  Note that this latter possesses a
companion lemma about inequalities, and both share the same proof through the
abstraction of their binary relation.

\subsection{{\FatouLem}}
\label{sec:bigth:fale_th}

{\FatouLem} (see Textbook Theorem~\ref{th:fatou}) is the other fundamental
result in Lebesgue integration theory for {\nonnegative} functions.  It
specifies how the situation deteriorates when the sequence in the
{\BeppoLeviTh} is no longer monotone: the equality becomes an inequality, and
limits ({\ie} suprema) become limits inferior.  It states that for any sequence
$(f_n)_{n\in\N}$ of {\nonnegative} measurable functions ({\ie} in~$\calMplus$),
the pointwise limit $\liminf_{n\to\infty}f_n$ is also in~$\calMplus$, and the
integral of the limit inferior is less than or equal to the limit inferior of
the integrals,
\[
  \intMplus \liminf_{n \to \infty} f_n \, d\mu
  \leq \liminf_{n \to \infty} \intMplus f_n \, d\mu,
\]
which is stated in {\Coq} as
\begin{lstlisting}
Theorem Fatou_lemma :
  \forall (f : nat -> E -> Rbar), (\forallp n, non_neg (f n)) ->
  (\forallp n, measurable_fun_Rbar genE (f n)) ->
    Rbar_le (LInt_p \musp (fun x => LimInf_seq' (fun n => f n x)))
             (LimInf_seq' (fun n => LInt_p \musp (f n))).
\end{lstlisting}

The proof is rather short.  The principle is to apply the {\BeppoLeviTh} (see
Section~\ref{sec:bigth:bele_tt}) to the sequence
$(\inf_{m\in\N}f_{n+m})_{n\in\N}$ which is shown {\nondecreasing},
{\nonnegative}, and measurable.  We get
\[
  \intMplus \liminf_{n \to \infty} f_n \, d\mu
  = \sup_{n \in \N} \intMplus \inf_{m \in \N} f_{n + m} d \mu.
\]
Then, by the {\monotony} result of Section~\ref{sec:lint_p_def} and the
definitions of infimum and limit inferior, we~have
\[
  \sup_{n \in \N} \intMplus \inf_{m \in \N} f_{n + m} d \mu
  \leq \liminf_{n \to \infty} \intMplus f_{n} \, d\mu,
\]
proving the result.

\myskip

Note that a less common proof path is also possible: establish first
{\FatouLem}, and then the {\BeppoLeviTh}.

{\FatouLem} is essential to establish other fundamental results such as the
{\FatouLebesgueTh} that collects a chain of inequalities involving both
inferior and superior limits, and above all, {\LebesgueDominatedConvergenceTh}
whose result is similar to that of the {\BeppoLeviTh}, but only through the
dominance of the sequence (both left as future work).

\section{A simple case study: the Dirac measure}
\label{sec:Dirac}

It makes sense to exhibit an example of measure to test the specifications
described in the previous sections, and especially the axiomatic definition of
Section~\ref{sec:measures}.

For instance, the Lebesgue measure, which extends the notion of length of
intervals in~$\R$, is ubiquitous on Euclidean spaces~$\R^n$.  And the counting
measure, which returns the cardinal, is pertinent on countable sets.  But both
present formalization issues, and their study is left for future works.

We present the construction and usage of a very simple measure, the Dirac
measure.  It is used for instance in physics to model a point mass.

\myskip

The Dirac map associated with an element $a\in E$ is the function~$\delta_a$
mapping any subset $A\subset E$ to~$\charac{A}(a)$ (see
Section~\ref{sec:charac}).
\begin{lstlisting}
Definition Dirac : E -> (E -> Prop) -> Rbar := fun a A => charac A a.
\end{lstlisting}
As a measure, the total function defined above only makes sense for measurable
subsets.  But~the Dirac measure has the salient property to be a measure for
any $\sigma$-algebra on~\coqe{E}, even for the discrete $\sigma$-algebra (see
Section~\ref{sec:sigma-alg}).

To instantiate the Dirac map~$\delta_a$ as a measure, we prove that it meets
the specification of measures of Section~\ref{sec:measures:spec_meas}.  The
first two properties, homogeneity (\coqe{Dirac_False}) and {\nonnegativeness}
(\coqe{Dirac_ge_0}), are obvious.  The proof of $\sigma$-addi\-tivity is based
on the following argument, that is valid for any pairwise disjoint sequence of
subsets $(A_n)_{n\in\N}$ (there is at most one {\nonzero} term in the sum).
\[
  \delta_a \left( \biguplus_{n \in \N} A_n \right)
  = \charac{\biguplus_{n \in \N} A_n} (a)
  = \sum_{n \in \N} \charac{A_n} (a)
  = \sum_{n \in \N} \delta_a (A_n).
\]
Then, the Dirac measure can be built for any generator \coqe{genE}.
\begin{lstlisting}
Definition Dirac_measure : E -> measure genE :=
  fun a => mk_measure genE (Dirac a)
    (Dirac_False a) (Dirac_ge_0 a) (Dirac_sigma_additivity a).
\end{lstlisting}

The integral of any function \coqe{f : E -> Rbar} with the Dirac measure is
\[
  \int f \, d\delta_a = f (a).
\]
The formalization in~$\calSFplus$ for {\nonnegative} measurable simple
functions, and any generator \coqe{genE}, is (see
Section~\ref{sec:lint_simple})
\begin{lstlisting}
Lemma LInt_SFp_Dirac :
  \forall (f : E -> R) (Hf : SF genE f) a, LInt_SFp (Dirac_measure genE a) f Hf = f a.
\end{lstlisting}
The proof is a direct application of lemma \coqe{finite_vals_sum_eq} of
Section~\ref{sec:canonizer}.  The version in~$\calMplus$ for {\nonnegative}
measurable functions, and any generator \coqe{genE}, is (see
Section~\ref{sec:lint_p_def})
\begin{lstlisting}
Lemma LInt_p_Dirac :
  \forall (f : E -> Rbar) a, non_neg f -> measurable_fun_Rbar genE f ->
    LInt_p E genE (Dirac_measure genE a) f = f a.
\end{lstlisting}
Its proof is an application of lemma \coqe{LInt_p_with_mk_adapted_seq} of
Section~\ref{sec:lint_p_adap_seq}.

\section{Discussion on proof-engineering concerns}
\label{sec:disc}

During our development, we had to make several choices regarding logic and the
formalization of~mathematics.

\subsection{Extended real numbers}
\label{sec:discuss:rbar}

Measure theory and integration of {\nonnegative} functions that are
investigated here only manipulate values in~$\Rbarplus$.  Thus arises the
question of the most practical {\Coq} formalization for {\nonnegative} extended
real numbers among the following three possible choices: either $[0,\infty]$,
$\R\cup\{\infty\}$, or $\R\cup\{-\infty,\infty\}$.

From a mathematical point of view, all solutions are acceptable because, in
addition, we either prove or require that values are {\nonnegative}.  But we
also need to keep in mind that eventually we will have to deal with arbitrary
sign functions.  And despite the absence of algebraic structure, extended real
numbers with both infinities are the usual framework often used by
mathematicians to allow for simplified expressions of many statements.  We have
chosen to follow this practice and to use \coqe{Rbar} from {\Coquelicot} (see
Section~\ref{sec:coquelicotRbar}).  But let us review the other possibilities
we considered.

\paragraph{First solution: $[0,\infty]$.}
For example with a specific type \coqe{Rbarplus}.  This would be very difficult
to use in {\Coq} because it would not be directly related to the type~\coqe{R},
so we would need a coercion or some subtyping to use this type in formulas with
reals.  Moreover, it would make~$\infty$ appear explicitly a lot.  This would
lead to very verbose statements with few automation possibilities.

\paragraph{Second solution: $\R\cup\{\infty\}$.}
For example with a type with two constructors~\coqe{R} and \coqe{p_infty}.
This~would keep validity of usual algebraic properties such as associativity
and distributivity without the need for additional hypotheses, and would favor
a low number of cases in proofs.  But~it would still be a new type that would
lead to {\Coq} coercions.  Moreover, when~$-\infty$ will enter the picture for
functions with arbitrary sign, it would make necessary coercions from/to the
three types, which would be as difficult to handle as coercions from/to~$\N$,
$\Z$, and~$\R$.

\paragraph{Chosen (third) solution: $\R\cup\{-\infty,\infty\}$.}
Which is the type \coqe{Rbar}.  This has the advantage of being already defined
in {\Coquelicot} with several lemmas proved, and is related to the
type~\coqe{R}.  Of~course, for the present developments on {\nonnegative}
functions, we have to deal with meaningless negative cases and additional
hypotheses.  However, this drawback is balanced by the fact that we are ready
to treat arbitrary sign functions.

\subsection{Classical logic aspects}
\label{sec:discuss:logic}

In our previous work, we tried to minimize the use of classical aspects.  For
example, in the formal proof of the {\LaxMilgramTh}~\cite{BCF17}, we had a few
decidability hypotheses and some statements relied on double negations to avoid
using a stronger classical property.  We consider here that it is no longer
worth the effort compared to the theoretical gain.  For this reason, we have
decided to use the classical theorems listed in Section~\ref{sec:basiccoq}.

Moreover, as explained in Section~\ref{sec:charac}, we choose that subsets have
the type \coqe{E -> Prop}.  We~could have chosen \coqe{E -> bool} and it would
have simplified some proofs as we rely a lot on the excluded middle axiom for
deciding whether a point is in a subset or not.  Nevertheless, this use of
\coqe{bool} would not have fully removed the excluded middle axiom.  Indeed,
when selecting in a list the elements that have a property (function
\coqe{select} of Section~\ref{sec:canonizer}), we need to decide inside this
function whether a property holds or not.  And this is then applied to the
following property: \coqe{fun x => \exists y, f y = x}.  We need, for each
$x\in E$ to decide whether it belongs to the image of~$f$ and that requires the
informative excluded middle axiom.  So, as it has no logical impact, we have
chosen \coqe{Prop} for convenience as it fits better in the libraries we rely
on.

\subsection{Measurability of subsets}
\label{sec:discuss:measurability}

Implementing the concept of (generated) $\sigma$-algebra, {\ie} the
measurability of subsets, as an inductive type allows to conduct proofs by
induction.  This is proposed in {\IsabelleHOL}\footnote{%
  The \coqe{sigma_sets} inductive set from
  \url{https://isabelle.in.tum.de/dist/library/HOL/HOL-Analysis/Sigma_Algebra.html}.}
and {\Lean},\footnote{%
  The \coqe{measurable_space.generate_measurable} inductive type from
  \url{https://leanprover-community.github.io/mathlib_docs/measure_theory/measurable_space.html\#measurable_space.generate_measurable}.}
and it is useful in {\Coq} too (see Section~\ref{sec:sigma-alg}).  But this
design choice can also have an impact on how mathematics results are stated.
Of course, there is a constructor for each basic property of $\sigma$-algebras.
But it is also necessary to add a constructor, \coqe{measurable_gen}, that
embodies the belonging to the collection of generators.  Indeed, this is
required to initiate the constructive process of specifying a measurable
subset.  In other words, our {\Coq} definition also encompasses the
mathematical concept of \emph{generated} $\sigma$-algebra.

As a consequence, we cannot instantiate a $\sigma$-algebra without exhibiting a
generator.  But~fortunately, nothing prevents from setting the whole
$\sigma$-algebra as the generator, and specify, or prove, that it satisfies
\coqe{is_sigma_algebra} predicate.  A notable effect is that the mathematical
result stating that any generated $\sigma$-algebra is the smallest
$\sigma$-algebra containing its generator is already structurally granted by
the inductive type \coqe{measurable}.  Our definition of generated
$\sigma$-algebra then precedes the definition of $\sigma$-algebra, which is not
the common order in mathematics.  It is however common to have several possible
orders or equivalent definitions in mathematics.  We found that this
formalization of generated $\sigma$-algebra was easy to use.  For example, the
{\Coq} proof of lemma \coqe{measurable_fun_gen} (see Section~\ref{sec:meas_fn})
is done very simply by induction on the measurable subset defined inductively
and applying directly the three properties \coqe{measurable_empty},
\coqe{measurable_compl} and \coqe{measurable_union_countable}.

\myskip

Whatever the definitions (or the order of them), the most difficult point was
related to the equivalence of generators.  More precisely, the equivalence
between the $\sigma$-algebra generated by compact intervals,
\coqe{measurable gen_R}, and the Borel $\sigma$-algebra generated by open
subsets, \coqe{measurable open}, of Section~\ref{sec:gen_equiv} has proved
tedious, long, and with harder ingredients than expected: many bijections,
connected components, density of~$\Q$, second-countability.

\subsection{Simple functions}
\label{sec:discuss:simplefun}

About simple functions, we had difficulties designing them and we tried many
{\Coq} definitions for the same mathematical object before deciding for the one
described in Section~\ref{sec:simple_fun}.

Note for instance that we have chosen the total function approach as much as
possible in our development to ensure the simplicity in writing formulas.  But
valid simple functions come as a dependent type~\coqe{SF} with the function,
the list of values and the corresponding proof.  This was needed as the list is
required to compute the integral.  To give the value of this integral, we need
to sum over a finite list and therefore we need this list to be given.  A
solution would be to have an easy mechanism to sum over arbitrary sets
(possibly infinite and possibly bigger than~$\N$) like done in
{\Lean}.\footnote{%
  The \coqe{tsum} operator from
  \url{https://leanprover-community.github.io/mathlib_docs/topology/algebra/infinite_sum.html\#tsum}.}
This extension of total function would make a practical addition to {\Coq} and
may simplify some of our statements but is out of the scope of this paper.
Note also that this dependent type~\coqe{SF} is not supposed to appear to a
library user, contrary to measurable functions and subsets, and to the
integral.

Another design choice is about the type of simple functions: either
\coqe{E -> R}, or \coqe{E -> Rbar}.  Mathematicians usually consider~$\R$ (in
fact~$\Rplus$) as codomain for~$\calSFplus$, and reserve~$\Rbar$ (in
fact~$\Rbarplus$) for~$\calMplus$ (as limits of functions in~$\calSFplus$).  We
first tried to have simple functions of type \coqe{E -> Rbar} for coherence and
simplicity, but it failed in the proof of the difficult
\coqe{sum_Rbar_map_change_of_variable} of
Section~\ref{sec:lint_simple_pos_lin}.  We could have kept the same type with
an assumption that all values taken are finite, but we found it less convenient
than using types for this requirement.  As suggested in mathematics, we ended
up by having simple functions of type \coqe{E -> R}.

A surprising successful choice is related to a particular type of simple
functions: adapted sequences.  Even if the mathematical definitions and proofs
are given, we chose to take a computer scientist (or even a computer
arithmetic) point of view.  The use of {\Flocq} has made trivial many~proofs.

\section{State of the Art}
\label{sec:soa}

Interactive theorem provers may be classified into several distinct families
with respect to their inherent logic.  It may rely on set theory or type
theory, on classical logic ({\eg} with a built-in axiom of choice) or
intuitionistic logic, and so on.  Similarly to programming languages, the
choice of a proof assistant is driven by the kind of formal proofs to be
developed, the support libraries, the ease of the developer, and so on.  It is
currently either impractical or impossible to automatically ``translate''
developments done within some proof assistant into another, especially when
they do not share the same inherent logic or theory.

This is the main reason for this work: providing to {\Coq} users a practical
formalization of Lebesgue integration.  As explained below, this already exists
in several other provers.  We have taken inspiration (within the limits set by
the various logics), we have made various design choices and proved the common
lemmas and theorems.  Our goal is to offer to {\Coq} users (so that they may
rely on it for their own {\Coq} development) this library.  This also allows
this library to be eventually mixed it with {\Flocq} for proving floating-point
programs.

\myskip

Lebesgue measure and integration is known to be an important chapter in
mathematics.  For~instance it belongs to a ``top 100'' of mathematical theorems
established at the turn of the millennium for which F.~Wiedijk is keeping track
of the formalization within the main proof assistants.\footnote{%
  \url{https://www.cs.ru.nl/~freek/100/}.}
This state of the art focuses on the formalization of the Lebesgue measure and
integration; for a larger view about real analysis (in {\ACLtwo}, {\Coq},
{\HOL}, {\HOLLight}, {\Mizar}, and {\PVS}), we refer the reader to this
survey~\cite{BLM16}.

\myskip

Regarding intrinsically classical proof assistants, we may cite
{\Mizar}~\cite{FormalMathsMizar}, {\IsabelleHOL}~\cite{Isabelle}, and
{\PVS}~\cite{PVS}.

{\Mizar} libraries have been very advanced since the 1990s on the formalization
of mathematics in general.  The Lebesgue integral has been addressed
continuously during the last two decades, for instance the integral of simple
functions~\cite{NoboruEndou2005}, {\FatouLem}~\cite{NoboruEndou2008b}, and
Fubini's theorem~\cite{NoboruEndou2019}.

More recently, a large library of results, including many results about measure
theory~\cite{HolHel11} and {\nonnegative} Lebesgue integration\footnote{%
  \url{https://isabelle.in.tum.de/dist/library/HOL/HOL-Analysis/Nonnegative_Lebesgue_Integration.html}.},
Ordinary Differential Equations (ODEs)~\cite{ImmlerH12,Immler14,ImmTrau16}, and
a formalization of {\GreenTh}~\cite{AbdLaw16} have been developed in
{\IsabelleHOL}, and Fourier transform~\cite{GuaZhaShiWanLi20} in {\HOLfour}.
In~{\HOLfour}~\cite{MhamdiHT10} and in {\PVS}\footnote{%
  \url{https://shemesh.larc.nasa.gov/fm/ftp/larc/PVS-library/library/measure_integration.html},
  \url{https://shemesh.larc.nasa.gov/fm/ftp/larc/PVS-library/library/lebesgue.html}.},
Lebesgue integration theory has also been developed in~2010.

In these provers, the definition of the Lebesgue integral is quite similar to
ours.  The main difference lies in the definition of simple functions.  Leaving
measurability aside, they consider the image set and the function is simple
when this set is finite.  For instance in PVS,
\texttt{is\_finite(image(f,\linebreak[0]%
  fullset[T]))}.

\myskip

Regarding intrinsically intuitionistic proof assistants, we may cite
{\Lean}~\cite{Lean}, and {\Coq}~\cite{Link_Coq_ref}.

About {\Lean}, we only found out in \cite{Leanmaths} that the Bochner integral
is available.  Note that the Bochner integral extends the Lebesgue integral to
functions taking their values in a Banach space.  There is no description on
how all this is formalized and the available theorems.  We investigated the
source, and found the {\BeppoLeviMonotConvTh} and {\FatouLem}, with some
formalization key points.  First, Borel spaces\footnote{%
  \url{https://leanprover-community.github.io/mathlib_docs/measure_theory/constructions/borel_space.html\#borel_space}.}
are generated from the open sets (only) with a kind of inductive definition, so
this ends up being similar to us, even if our genericity in terms of generators
has proved useful.  Second, as in other provers, simple function
definitions\footnote{%
  \url{https://leanprover-community.github.io/mathlib_docs/measure_theory/integral/lebesgue.html\#measure_theory.simple_func}.}
rely on a predicate that says whether a set is finite or not (used on the image
of the function), which relies on the logical underlying framework that is
quite different from the one of {\Coq}.

As mentioned in the introduction, our work relies on the {\Coq} proof
assistant.  As a previous work about analysis in {\Coq} we can cite
formalization of Picard's operator for ODEs~\cite{MakSpi13}, which uses the
constructive {\CoRN} and the {\MathClasses} libraries.  Regarding analysis with
classical reals, our previous works in {\Coq} include the full formalization of
the discretization of the wave equation~\cite{BCF13,bol:tcm:14}, and the
formalization of {\LaxMilgramTh}~\cite{BCF17}.  For both of these works, we
paid particular attention to the statements and their proof to avoid the use of
classical axioms (we have now chosen the classical side, see
Section~\ref{sec:discuss:logic}).  Another recent development is the formal
proof of the Lax equivalence theorem for finite difference
schemes~\cite{Tekriwal21}, it is based on the classical standard real numbers,
{\Coquelicot} libraries, and our formalization of the {\LaxMilgramTh}.

On the other hand, the {\MathCompAnalysis} library is currently being
developed~\cite{ACMRS18}.  It aims at providing numerical analysis results in
classical logic, building upon {\MathComp}.  This library has been developed in
parallel to ours and is still in development, with few documentation and many
branches, so it is hard to trace proved theorems.\footnote{%
  \url{https://github.com/math-comp/analysis}.}
We found out similar definitions for $\sigma$-algebra (by induction), connected
components and the Lebesgue integral.  In one branch, we found out a very
experimental generic integral but in another one dedicated to the Lebesgue
integral, we found some of the lemmas described in this article.  As for the
differences, they define simple functions by a dependent type including a
unique list.  Our adapted sequences are more smooth as based on {\Flocq}, while
they handle dyadic intervals by hand.  They have a full definition of the
Lebesgue measure using {\CaratheodoryTh}.

Last, following the rebuilding of the standard real library~\cite{Semeria20},
in which a constructive and classical basis was built up, constructive analysis
lemmas were also introduced by Vincent Semeria, based on the constructive
analysis of Bishop~\cite{bis:fca:67}.

\mybreak
\section{Conclusion and perspectives}
\label{sec:concl}

This work is a second stone that paves the way toward the formal correctness of
the {\FEM}, the first one being the formal proof of the
{\LaxMilgramTh}~\cite{BCF17}.  The~contributions are the {\Coq} formalizations
and proofs of $\sigma$-algebras, measures, simple functions, and Lebesgue
integration of {\nonnegative} measurable functions, and the formal proofs of
the {\BeppoLeviMonotConvTh} and {\FatouLem}.

The subset addressed in the present paper is more than 50-page long (6000 lines
of code (LOC) of {\LaTeX}, and weighs 220~kB) in its mathematics
counterpart~\cite{cm:li:21}.  These {\Coq} source files add up to about
11~kLOC, and weigh 370~kB.  In total, the cumulative development including the
{\LaxMilgramTh}~\cite{BCF17}, finite-dimensional vector
spaces~\cite{Faissole2017formalization}, and this formalization is about
21~kLOC/650~kB.

As in~\cite{bol:tcm:14}, we observe here again that in-depth pen-and-paper
proofs can be an order of magnitude longer than usual proofs from textbooks,
and the lengths of formal and detailed pen-and-paper proofs are similar.  In
the present case, we can notice that the extra effort deployed on \coqe{Rbar}
(in \coqe{Rbar_compl} and \coqe{sum_Rbar_nonneg}, roughly 3~kLOC/90~kB)
explains part of the gap between the respective sizes of the {\Coq}
formalization and the pen-and-paper proof, where~$\Rbar$ received far less
detail.

\begin{figure}[ht]
  \centering
  \input{depend_txt}
  \caption{%
    Dependency graph of our {\Coq} development.\protect\\
    {\ComplColorName}: complements to standard libraries and
    {\Coquelicot}.\protect\\
    {\PrelimColorName}: new preliminary developments.\protect\\
    {\MeasColorName}: new developments in measure theory.\protect\\
    {\LintColorName}: new developments in Lebesgue integration.}
  \label{fig:dep}
\end{figure}

The hyperlinked dependency graph is detailed in Figure~\ref{fig:dep} where our
target development, the Lebesgue integral (built on simple functions), is
represented in {\lintColorName}.  For this purpose, we had to formalize
$\sigma$-algebras, measurable functions, and measures (represented in
{\measColorName}), as well as some preliminary developments on countability,
topological bases in~$\R$, and the handling of sums in~$\Rbar$ (represented in
{\prelimColorName}).  We also had to develop results that were missing both in
the standard libraries (in the subdirectories {\Logic}, {\Lists}, {\Sorting},
and {\Reals}) and {\Coquelicot} (represented in {\complColorName}), including
some tactics for~\coqe{Rbar}.  As usual, we can note that a large number of
prerequisites are necessary to reach the desired formalization.  In our case,
this distributes roughly into one third for the \complColor{complements}
(either in~kLOC or in~kB), one seventh for the \prelimColor{preliminaries}, one
third for \measColor{measure theory}, and one fifth for the target
\lintColor{Lebesgue integral}.

As usual, formalization is not just straightforward translation of mathematical
texts and formulas.  Some design choices have to be made and proof paths may
differ, mainly to favor usability of {\Coq} theorems and ease formal
developments.

\myskip

After both the {\LaxMilgramTh}~\cite{BCF17} and this work, the road is still
long to be able to tackle the formal proof of scientific computation programs
using the {\FEM} (FEM).

The next step is to formalize the construction of the Lebesgue measure (for
instance using {\CaratheodoryTh}~\cite{car:atm:63,dur:pte:19}), and the
Lebesgue integral for measurable functions with arbitrary sign, with the proofs
of {\LebesgueDominatedConvergenceTh} and of the {\TonelliFubiniThs} as the next
milestones.  Then comes the formalization of the~$L^p$ Lebesgue spaces as
complete normed vector spaces ({\aka} Banach spaces), and in particular
of~$L^2$ as a complete inner product space ({\aka} a Hilbert space).  We expect
the completeness to be the most challenging part of the proof.  And finally,
the formalization of simple Sobolev spaces~$H^1$ and~$H^1_0$ will need part of
the distribution theory~\cite{sch:td:66}.  Further steps will include the
formalization of parts of interpolation and approximation theories to end up
with the FEM.

In parallel, we plan to merge with recent works on constructive
reals~\cite{Semeria20} now included in the {\Coq} standard library, and in
particular with the constructive measure theory~\cite{bc:cmt:72} based on the
Daniell integral~\cite{dan:gfi:18}.  We also plan to formalize the Bochner
integral~\cite{boc:ifw:33,mik:bi:78} that generalizes the Lebesgue integral to
the case of functions taking their values in a Banach space, for instance such
as the Euclidean spaces~$\R^n$ and the Hermitian spaces~$\C^n$.

\section*{Acknowledgments}

We are deeply grateful to Stéphane Aubry for some proofs and tactics on
\coqe{Rbar} properties, and to Assia Mahboubi, Cyril Cohen, Guillaume
Melquiond, and Vincent Semeria for fruitful discussions about \coqe{Rbar} and
\coqe{R} in {\Coq}.

This work was partly supported by the Paris Île-de-France Region (DIM RFSI
MILC).

This work was partly supported by Labex DigiCosme (project
ANR-11-LABEX-0045-DIGI\-COSME) operated by ANR as part of the program
``Investissement d'Avenir'' Idex Paris-Saclay (ANR-11-IDEX-0003-02).

This work was partly supported by the European Research Council (ERC) under the
European Union's Horizon 2020 Research and Innovation Programme – Grant
Agreement n$^\circ$810367.

\bibliographystyle{plainnat}
\bibliography{biblio}

\begin{thebibliography}{70}
\providecommand{\natexlab}[1]{#1}
\providecommand{\url}[1]{\texttt{#1}}
\expandafter\ifx\csname urlstyle\endcsname\relax
  \providecommand{\doi}[1]{doi: #1}\else
  \providecommand{\doi}{doi: \begingroup \urlstyle{rm}\Url}\fi

\bibitem[Abdulaziz and Paulson(2016)]{AbdLaw16}
Mohammad Abdulaziz and Lawrence~C. Paulson.
\newblock An {I}sabelle/{HOL} formalisation of {G}reen's theorem.
\newblock In Christian~Jasmin Blanchette and Stephan Merz, editors, \emph{Proc.
  of the 7th Internat. Conf. on Interactive Theorem Proving ({ITP 2016})},
  volume 9807 of \emph{Lecture Notes in Computer Science}, pages 3--19.
  Springer, Cham, 2016.
\newblock URL \url{https://doi.org/10.1007/978-3-319-43144-4_1}.

\bibitem[Adams(1975)]{ada:ss:75}
Robert~A. Adams.
\newblock \emph{Sobolev Spaces}, volume~65 of \emph{Pure and Applied
  Mathematics}.
\newblock Academic Press, New York - San Francisco - London, 1975.

\bibitem[Affeldt et~al.(2018)Affeldt, Cohen, Mahboubi, Rouhling, and
  Strub]{ACMRS18}
Reynald Affeldt, Cyril Cohen, Assia Mahboubi, Damien Rouhling, and Pierre-Yves
  Strub.
\newblock Classical analysis with {C}oq.
\newblock The 9th Coq Workshop, 2018.
\newblock URL \url{https://easychair.org/smart-slide/slide/n3pK}.

\bibitem[Bartle(2001)]{bar:mti:01}
Robert~G. Bartle.
\newblock \emph{A Modern Theory of Integration}, volume~32 of \emph{Graduate
  Studies in Mathematics}.
\newblock American Mathematical Society, Providence, 2001.
\newblock URL \url{https://doi.org/10.1090/gsm/032}.

\bibitem[Bertot et~al.(2008)Bertot, Gonthier, Biha, and
  Pasca]{Bertot2008canonical}
Yves Bertot, Georges Gonthier, Sidi~Ould Biha, and Ioana Pasca.
\newblock Canonical big operators.
\newblock In Otmane~Ait Mohamed, César Muñoz, and Sofiène Tahar, editors,
  \emph{Proc. of the 21st Internat. Conf. on Theorem Proving in Higher Order
  Logics ({TPHOL 2008})}, volume 5170 of \emph{Lecture Notes in Computer
  Science}, pages 86--101. Springer, Berlin - Heidelberg, 2008.
\newblock URL \url{https://doi.org/10.1007/978-3-540-71067-7_11}.

\bibitem[Billingsley(1995)]{bil:pm:95}
Patrick Billingsley.
\newblock \emph{Probability and Measure}.
\newblock Wiley Series in Probability and Mathematical Statistics. John Wiley
  \& Sons, Inc., New York, 3rd edition, 1995.

\bibitem[Bishop(1967)]{bis:fca:67}
Errett Bishop.
\newblock \emph{Foundations of Constructive Analysis}.
\newblock McGraw-Hill Book Co., New York - Toronto - London, 1967.

\bibitem[Bishop and Cheng(1972)]{bc:cmt:72}
Errett Bishop and Henry Cheng.
\newblock \emph{Constructive Measure Theory}.
\newblock Number 116 in Memoirs of the American Mathematical Society. American
  Mathematical Society, Providence, 1972.

\bibitem[Bochner(1933)]{boc:ifw:33}
Salomon Bochner.
\newblock Integration von funktionen, deren werte die elemente eines
  vektorraumes sind.
\newblock \emph{Fundam. Math.}, 20:\penalty0 262--276, 1933.
\newblock In German.

\bibitem[Boldo and Melquiond(2011)]{BolMel11}
Sylvie Boldo and Guillaume Melquiond.
\newblock {Flocq: A Unified Library for Proving Floating-point Algorithms in
  Coq}.
\newblock In \emph{Proc. of the IEEE 20th Symposium on Computer Arithmetic
  ({ARITH-20})}, pages 243--252. IEEE, 2011.
\newblock URL \url{https://doi.org/10.1109/ARITH17396.2011}.

\bibitem[Boldo et~al.(2013)Boldo, Clément, Filliâtre, Mayero, Melquiond, and
  Weis]{BCF13}
Sylvie Boldo, François Clément, Jean-Christophe Filliâtre, Micaela Mayero,
  Guillaume Melquiond, and Pierre Weis.
\newblock {Wave Equation Numerical Resolution: a Comprehensive Mechanized Proof
  of a {C} Program}.
\newblock \emph{J. Autom. Reason.}, 50\penalty0 (4):\penalty0 423--456, 2013.
\newblock URL \url{https://hal.inria.fr/hal-00649240/}.

\bibitem[Boldo et~al.(2014)Boldo, Clément, Filliâtre, Mayero, Melquiond, and
  Weis]{bol:tcm:14}
Sylvie Boldo, François Clément, Jean-Christophe Filliâtre, Micaela Mayero,
  Guillaume Melquiond, and Pierre Weis.
\newblock Trusting computations: a mechanized proof from partial differential
  equations to actual program.
\newblock \emph{Comput. Math. with Appl.}, 68\penalty0 (3):\penalty0 325--352,
  2014.
\newblock URL \url{https://hal.inria.fr/hal-00769201/}.

\bibitem[Boldo et~al.(2015)Boldo, Lelay, and Melquiond]{BLM15}
Sylvie Boldo, Catherine Lelay, and Guillaume Melquiond.
\newblock Coquelicot: A user-friendly library of real analysis for {Coq}.
\newblock \emph{Math. Comput. Sci.}, 9\penalty0 (1):\penalty0 41--62, 2015.
\newblock URL \url{https://hal.inria.fr/hal-00860648/}.

\bibitem[Boldo et~al.(2016)Boldo, Lelay, and Melquiond]{BLM16}
Sylvie Boldo, Catherine Lelay, and Guillaume Melquiond.
\newblock {Formalization of Real Analysis: A Survey of Proof Assistants and
  Libraries}.
\newblock \emph{Math. Struct. Comput. Sci.}, 26\penalty0 (7):\penalty0
  1196--1233, 2016.
\newblock URL \url{https://hal.inria.fr/hal-00806920/}.

\bibitem[Boldo et~al.(2017)Boldo, Clément, Faissole, Martin, and
  Mayero]{BCF17}
Sylvie Boldo, François Clément, Florian Faissole, Vincent Martin, and Micaela
  Mayero.
\newblock A {C}oq formal proof of the {L}ax--{M}ilgram theorem.
\newblock In \emph{Proc. of the 6th ACM SIGPLAN Internat. Conf. on Certified
  Programs and Proofs ({CPP 2017})}, pages 79--89. Association for Computing
  Machinery, New York, 2017.
\newblock URL \url{https://hal.inria.fr/hal-01391578/}.

\bibitem[Bourbaki(1965)]{bou:int:65}
Nicolas Bourbaki.
\newblock \emph{Éléments de mathématiques. Livre~VI~: Intégration.
  Chapitres~1 à~4}.
\newblock Hermann, Paris, 2nd edition, 1965.
\newblock In French.

\bibitem[Bourbaki(1971)]{bou:tg:71}
Nicolas Bourbaki.
\newblock \emph{Éléments de mathématiques. Livre~III~: Topologie générale.
  Chapitres~1 à~4}.
\newblock Hermann, Paris, 2nd edition, 1971.
\newblock In French.

\bibitem[Brezis(1983)]{bre:af:83}
Haïm Brezis.
\newblock \emph{Analyse fonctionnelle---Théorie et applications}.
\newblock Collection Mathématiques Appliquées pour la Maîtrise. Masson,
  Paris, 1983.
\newblock In French.

\bibitem[Burk(2007)]{bur:gi:07}
Frank~E. Burk.
\newblock \emph{A Garden of Integrals}, volume~31 of \emph{The Dolciani
  Mathematical Expositions}.
\newblock Mathematical Association of America, Washington, 2007.

\bibitem[Carathéodory(1963)]{car:atm:63}
Constantin Carathéodory.
\newblock \emph{Algebraic Theory of Measure and Integration}.
\newblock Chelsea Publishing Co., New York, 1963.

\bibitem[Cartan(1937)]{car:tf:37}
Henri Cartan.
\newblock Théorie des filtres.
\newblock \emph{C. R. Acad. Sci.}, 205:\penalty0 595--598, 1937.
\newblock In French.

\bibitem[Ciarlet(2002)]{cia:fem:02}
Philippe~G. Ciarlet.
\newblock \emph{The Finite Element Method for Elliptic Problems}, volume~40 of
  \emph{Classics in Applied Mathematics}.
\newblock Society for Industrial and Applied Mathematics (SIAM), Philadelphia,
  2002.
\newblock URL \url{https://doi.org/10.1137/1.9780898719208}.
\newblock Reprint of the 1978 original [North-Holland, Amsterdam].

\bibitem[Clément and Martin(2021)]{cm:li:21}
François Clément and Vincent Martin.
\newblock {L}ebesgue integration. {D}etailed proofs to be formalized in {C}oq.
\newblock Research Report RR-9386, Inria, Paris, Jan 2021.
\newblock URL \url{https://hal.inria.fr/hal-03105815v2}.
\newblock Version 2.

\bibitem[Cohn(2013)]{coh:mt:13}
Donald~L. Cohn.
\newblock \emph{Measure Theory}.
\newblock Birkhäuser Advanced Texts: Basler Lehrbücher. Birkhäuser/Springer,
  New York, 2nd edition, 2013.
\newblock URL \url{https://doi.org/10.1007/978-1-4614-6956-8}.

\bibitem[Coq-ref()]{Link_Coq_ref}
Coq-ref.
\newblock The {Coq} reference manual.
\newblock URL \url{https://coq.inria.fr/refman/}.

\bibitem[Daniell(1918)]{dan:gfi:18}
Percy~John Daniell.
\newblock A general form of integral.
\newblock \emph{Ann. Math. (2)}, 19\penalty0 (4):\penalty0 279--294, 1918.
\newblock URL \url{https://doi.org/10.2307/1967495}.

\bibitem[de~Moura et~al.(2015)de~Moura, Kong, Avigad, van Doorn, and von
  Raumer]{Lean}
Leonardo de~Moura, Soonho Kong, Jeremy Avigad, Floris van Doorn, and Jakob von
  Raumer.
\newblock The {L}ean theorem prover (system description).
\newblock In Amy~P. Felty and Aart Middeldorp, editors, \emph{Proc. of the 25th
  Internat. Conf. on Automated Deduction ({CADE 2015})}, volume 9195 of
  \emph{Lecture Notes in Computer Science}, pages 378--388. Springer, Cham,
  2015.
\newblock URL \url{https://doi.org/10.1007/978-3-319-21401-6_26}.

\bibitem[Dieudonné(1968)]{die:ea2:68}
Jean Dieudonné.
\newblock \emph{Éléments d'analyse. Tome II~: Chapitres~XII à~XV}.
\newblock Cahiers Scientifiques, Fasc. XXXI. Gauthier-Villars, Paris, 1968.
\newblock In French.

\bibitem[Durrett(2019)]{dur:pte:19}
Richard Durrett.
\newblock \emph{Probability---Theory and Examples}, volume~49 of
  \emph{Cambridge Series in Statistical and Probabilistic Mathematics}.
\newblock Cambridge University Press, Cambridge, 5th edition, 2019.
\newblock URL \url{https://doi.org/10.1017/9781108591034}.

\bibitem[Endou(2019)]{NoboruEndou2019}
Noboru Endou.
\newblock Fubini's theorem.
\newblock \emph{Formaliz. Math.}, 27\penalty0 (1):\penalty0 67--74, 2019.
\newblock URL \url{https://doi.org/10.2478/forma-2019-0007}.

\bibitem[Endou et~al.(2005)Endou, Narita, and Shidama]{NoboruEndou2005}
Noboru Endou, Keiko Narita, and Yasunari Shidama.
\newblock Lebesgue integral of simple valued function in {M}izar.
\newblock \emph{Formaliz. Math.}, 13\penalty0 (1):\penalty0 67--71, 2005.
\newblock URL \url{https://fm.mizar.org/2005-13/pdf13-1/mesfunc3.pdf}.

\bibitem[Endou et~al.(2008)Endou, Narita, and Shidama]{NoboruEndou2008b}
Noboru Endou, Keiko Narita, and Yasunari Shidama.
\newblock Fatou's lemma and the {L}ebesgue's convergence theorem.
\newblock \emph{Formaliz. Math.}, 16\penalty0 (4):\penalty0 305--309, 2008.
\newblock URL \url{https://doi.org/10.2478/v10037-008-0037-8}.

\bibitem[Ern and Guermond(2004)]{eg:tpf:04}
Alexandre Ern and Jean-Luc Guermond.
\newblock \emph{Theory and Practice of Finite Elements}, volume 159 of
  \emph{Applied Mathematical Sciences}.
\newblock Springer, New York, 2004.
\newblock URL \url{https://doi.org/10.1007/978-1-4757-4355-5}.

\bibitem[Faissole(2017)]{Faissole2017formalization}
Florian Faissole.
\newblock Formalization and closedness of finite dimensional subspaces.
\newblock In \emph{Proc. of the 19th Internat. Symposium on Symbolic and
  Numeric Algorithms for Scientific Computing ({SYNASC 2017})}, pages 121--128.
  IEEE, 2017.

\bibitem[Feller(1968)]{fel:ipt:68}
William Feller.
\newblock \emph{An Introduction to Probability Theory and its Applications.
  {V}ol.~{I}}.
\newblock John Wiley \& Sons, Inc., New York - London - Sydney, 3rd edition,
  1968.

\bibitem[Folland(1999)]{fol:ram:99}
Gerald~B. Folland.
\newblock \emph{Real Analysis---Modern Techniques and Their Applications}.
\newblock Pure and Applied Mathematics (New York). John Wiley \& Sons, Inc.,
  New York, 2nd edition, 1999.

\bibitem[Gallouët and Herbin(2013)]{gh:mip:13}
Thierry Gallouët and Raphaèle Herbin.
\newblock \emph{Mesure, intégration, probabilités}.
\newblock Ellipses Edition Marketing, 2013.
\newblock URL \url{https://hal.archives-ouvertes.fr/hal-01283567/}.
\newblock In French.

\bibitem[Ghosal and van~der Vaart(2017)]{gv:fnb:17}
Subhashis Ghosal and Aad van~der Vaart.
\newblock \emph{Fundamentals of Nonparametric {B}ayesian Inference}, volume~44
  of \emph{Cambridge Series in Statistical and Probabilistic Mathematics}.
\newblock Cambridge University Press, Cambridge, 2017.
\newblock URL \url{https://doi.org/10.1017/9781139029834}.

\bibitem[Gill and Zachary(2016)]{gw:faf:16}
Tepper~L. Gill and Woodford~W. Zachary.
\newblock \emph{Functional Analysis and the {F}eynman Operator Calculus}.
\newblock Springer, Cham, 2016.
\newblock URL \url{https://doi.org/10.1007/978-3-319-27595-6}.

\bibitem[Gostiaux(1993)]{gos:cms2:93}
Bernard Gostiaux.
\newblock \emph{Cours de mathématiques spéciales - 2.~Topologie, analyse
  réelle}.
\newblock Mathématiques. Presses Universitaires de France, Paris, 1993.
\newblock In French.

\bibitem[Guan et~al.(2020)Guan, Zhang, Shi, Wang, and Li]{GuaZhaShiWanLi20}
Yong Guan, Jie Zhang, Zhiping Shi, Yi~Wang, and Yongdong Li.
\newblock Formalization of continuous {F}ourier transform in verifying
  applications for dependable cyber-physical systems.
\newblock \emph{J. Syst. Archit.}, 106, 2020.
\newblock URL \url{https://doi.org/10.1016/j.sysarc.2020.101707}.

\bibitem[Henstock(1963)]{hen:ti:63}
Ralph Henstock.
\newblock \emph{Theory of Integration}.
\newblock Buttherworths, London, 1963.

\bibitem[Hölzl and Heller(2011)]{HolHel11}
Johannes Hölzl and Armin Heller.
\newblock Three chapters of measure theory in {Isabelle/HOL}.
\newblock In Marko van Eekelen, Herman Geuvers, Julien Schmaltz, and Freek
  Wiedijk, editors, \emph{Proc. of the 2nd Internat. Conf. on Interactive
  Theorem Proving ({ITP 2011})}, volume 6898 of \emph{Lecture Notes in Computer
  Science}, pages 135--151. Springer, Berlin - Heidelberg, 2011.
\newblock URL \url{https://doi.org/10.1007/978-3-642-22863-6_12}.

\bibitem[Immler(2014)]{Immler14}
Fabian Immler.
\newblock Formally verified computation of enclosures of solutions of ordinary
  differential equations.
\newblock In Julia~M. Badger and Kristin~Yvonne Rozier, editors, \emph{Proc. of
  the 6th Internat. Symp. {NASA} Formal Methods ({NFM 2014})}, volume 8430 of
  \emph{Lecture Notes in Computer Science}, pages 113--127. Springer, Cham,
  2014.
\newblock URL \url{https://doi.org/10.1007/978-3-319-06200-6_9}.

\bibitem[Immler and Hölzl(2012)]{ImmlerH12}
Fabian Immler and Johannes Hölzl.
\newblock Numerical analysis of ordinary differential equations in
  {I}sabelle/{HOL}.
\newblock In Lennart Beringer and Amy~P. Felty, editors, \emph{Proc. of the 3rd
  Internat. Conf. on Interactive Theorem Proving ({ITP 2012})}, volume 7406 of
  \emph{Lecture Notes in Computer Science}, pages 377--392. Springer, Berlin -
  Heidelberg, 2012.
\newblock URL \url{https://doi.org/10.1007/978-3-642-32347-8_26}.

\bibitem[Immler and Traut(2016)]{ImmTrau16}
Fabian Immler and Christoph Traut.
\newblock The flow of {ODEs}.
\newblock In Christian~Jasmin Blanchette and Stephan Merz, editors, \emph{Proc.
  of the 7th Internat. Conf. on Interactive Theorem Proving ({ITP 2016})},
  volume 9807 of \emph{Lecture Notes in Computer Science}, pages 184--199.
  Springer, Cham, 2016.
\newblock URL \url{https://doi.org/10.1007/978-3-319-43144-4_12}.

\bibitem[Kurzweil(1957)]{kur:god:57}
Jaroslav Kurzweil.
\newblock Generalized ordinary differential equations and continuous dependence
  on a parameter.
\newblock \emph{Czechoslov. Math. J.}, 7\penalty0 (3):\penalty0 418--449, 1957.
\newblock URL \url{https://doi.org/10.21136/CMJ.1957.100258}.

\bibitem[Lebesgue(2009)]{leb:lir:04}
Henri~Léon Lebesgue.
\newblock \emph{Leçons sur l'intégration et la recherche des fonctions
  primitives professées au {C}ollège de {F}rance}.
\newblock Cambridge Library Collection. Cambridge University Press, Cambridge,
  2009.
\newblock URL \url{https://doi.org/10.1017/CBO9780511701825}.
\newblock Reprint of the 1904 original [Gauthier-Villars, Paris]. In French.

\bibitem[Lelay(2015{\natexlab{a}})]{Lelay15}
Catherine Lelay.
\newblock \emph{Repenser la bibliothèque réelle de {Coq} : vers une
  formalisation de l'analyse classique mieux adaptée}.
\newblock Thèse de doctorat, Université Paris-Sud, June 2015{\natexlab{a}}.
\newblock URL \url{https://tel.archives-ouvertes.fr/tel-01228517/}.
\newblock In French.

\bibitem[Lelay(2015{\natexlab{b}})]{Lelay15coq}
Catherine Lelay.
\newblock How to express convergence for analysis in {Coq}.
\newblock In \emph{The 7th Coq Workshop}, June 2015{\natexlab{b}}.
\newblock URL \url{https://hal.archives-ouvertes.fr/hal-01169321/}.

\bibitem[Maisonneuve(2014)]{mai:m2:14}
Francis Maisonneuve.
\newblock \emph{Mathématiques~2~: Intégration, transformations, intégrales
  et applications - Cours et exercices}.
\newblock Presses de l'École des Mines, 2014.
\newblock In French.

\bibitem[Makarov and Spitters(2013)]{MakSpi13}
Evgeny Makarov and Bas Spitters.
\newblock The {P}icard algorithm for ordinary differential equations in {C}oq.
\newblock In Sandrine Blazy, Christine Paulin-Mohring, and David Pichardie,
  editors, \emph{Proc. of the 4th Internat. Conf. on Interactive Theorem
  Proving ({ITP 2013})}, volume 7998 of \emph{Lecture Notes in Computer
  Science}, pages 463--468. Springer, Berlin - Heidelberg, 2013.
\newblock URL \url{https://doi.org/10.1007/978-3-642-39634-2_34}.

\bibitem[Matuszewski()]{FormalMathsMizar}
Roman Matuszewski, editor.
\newblock \emph{Formalized Mathematics}.
\newblock Sciendo, Poland.
\newblock URL \url{https://fm.mizar.org/}.

\bibitem[Mayero(2001)]{May01}
Micaela Mayero.
\newblock \emph{Formalisation et automatisation de preuves en analyses réelle
  et numérique}.
\newblock Thèse de doctorat, Université Paris VI, December 2001.
\newblock URL
  \url{http://www-lipn.univ-paris13.fr/~mayero/publis/these-mayero.ps.gz}.
\newblock In French.

\bibitem[Mhamdi et~al.(2010)Mhamdi, Hasan, and Tahar]{MhamdiHT10}
Tarek Mhamdi, Osman Hasan, and Sofiène Tahar.
\newblock On the formalization of the lebesgue integration theory in {HOL}.
\newblock In Matt Kaufmann and Lawrence~C. Paulson, editors, \emph{Proc. of the
  1st Internat. Conf. on Interactive Theorem Proving ({ITP 2010})}, volume 6172
  of \emph{Lecture Notes in Computer Science}, pages 387--402. Springer, Berlin
  - Heidelberg, 2010.
\newblock URL \url{https://doi.org/10.1007/978-3-642-14052-5_27}.

\bibitem[Mikusi\'nski(1978)]{mik:bi:78}
Jan Mikusi\'nski.
\newblock \emph{The Bochner Integral}.
\newblock Academic Press, New York - San Francisco, 1978.

\bibitem[Musial and Tulone(2019)]{mt:dch:19}
Paul Musial and Francesco Tulone.
\newblock Dual of the class of {HK}$_r$ integrable functions.
\newblock \emph{Minimax Theory its Appl.}, 4\penalty0 (1):\penalty0 151--160,
  2019.

\bibitem[Nipkow et~al.(2002)Nipkow, Paulson, and Wenzel]{Isabelle}
Tobias Nipkow, Lawrence~C. Paulson, and Markus Wenzel.
\newblock \emph{Isabelle/HOL---A Proof Assistant for Higher-Order Logic},
  volume 2283 of \emph{Lecture Notes in Computer Science}.
\newblock Springer, Berlin - Heidelberg - New York, 2002.
\newblock URL \url{https://doi.org/10.1007/3-540-45949-9}.

\bibitem[Owre et~al.(1992)Owre, Rushby, and Shankar]{PVS}
Sam Owre, John~M. Rushby, and Natarajan Shankar.
\newblock {PVS: A} prototype verification system.
\newblock In Deepak Kapur, editor, \emph{Proc. of the 11th Internat. Conf. on
  Automated Deduction ({CADE 1992})}, volume 607 of \emph{Lecture Notes in
  Computer Science}, pages 748--752. Springer, Berlin - Heidelberg, 1992.
\newblock URL \url{https://doi.org/10.1007/3-540-55602-8_217}.

\bibitem[Owre et~al.(2020)Owre, Shankar, Rushby, and
  Stringer-Calvert]{Owre1999pvs:alt}
Sam Owre, Natarajan Shankar, John~M. Rushby, and David~W.J. Stringer-Calvert.
\newblock \emph{PVS System Guide}.
\newblock SRI International, Computer Science Laboratory, Menlo Park, CA,
  August 2020.
\newblock URL \url{http://pvs.csl.sri.com/doc/pvs-system-guide.pdf}.
\newblock Version 7.1 [1st version in 1999].

\bibitem[Quarteroni and Valli(1994)]{qv:nap:94}
Alfio Quarteroni and Alberto Valli.
\newblock \emph{Numerical approximation of partial differential equations},
  volume~23 of \emph{Springer Series in Computational Mathematics}.
\newblock Springer, Berlin, 1994.

\bibitem[Rudin(1987)]{rud:rca:87}
Walter Rudin.
\newblock \emph{Real and Complex Analysis}.
\newblock McGraw-Hill Book Co., New York, 3rd edition, 1987.

\bibitem[Schwartz(1966)]{sch:td:66}
Laurent Schwartz.
\newblock \emph{Théorie des distributions}.
\newblock Hermann, Paris, 2nd edition, 1966.
\newblock 1st edition in 1950--1951. In French.

\bibitem[Semeria(2020)]{Semeria20}
Vincent Semeria.
\newblock Nombres réels dans {C}oq.
\newblock In Zaynah Dargaye and Yann Regis-Gianas, editors, \emph{Actes des
  31es Journées Francophones des Langages Applicatifs (JFLA)}, pages 104--111.
  IRIF, 2020.
\newblock URL \url{https://hal.inria.fr/hal-02427360/}.
\newblock In French.

\bibitem[Tekriwal et~al.(2021)Tekriwal, Duraisamy, and Jeannin]{Tekriwal21}
Mohit Tekriwal, Karthik Duraisamy, and Jean-Baptiste Jeannin.
\newblock A formal proof of the {L}ax equivalence theorem for finite difference
  schemes.
\newblock In \emph{13th Internat. Symp. {NASA} Formal Methods ({NFM 2021})},
  2021.
\newblock To appear.

\bibitem[{The mathlib Community}(2020)]{Leanmaths}
{The mathlib Community}.
\newblock The {L}ean mathematical library.
\newblock In Jasmin Blanchette and Catalin Hritcu, editors, \emph{Proc. of the
  9th {ACM} {SIGPLAN} Internat. Conf. on Certified Programs and Proofs ({CPP
  2020})}, pages 367--381. {ACM}, 2020.
\newblock URL \url{https://doi.org/10.1145/3372885.3373824}.

\bibitem[Tsybakov(2009)]{tsy:ine:09}
Alexandre~B. Tsybakov.
\newblock \emph{Introduction to Nonparametric Estimation}.
\newblock Springer Series in Statistics. Springer, New York, 2009.
\newblock URL \url{https://doi.org/10.1007/b13794}.
\newblock Revised and extended from the 2004 French original [Springer,
  Berlin], translated by Vladimir Zaiats.

\bibitem[Weil(1937)]{wei:esu:37}
André Weil.
\newblock \emph{Sur les espaces à structure uniforme et sur la topologie
  générale}.
\newblock Hermann, Paris, 1937.
\newblock In French.

\bibitem[Yosida(1995)]{yos:fa:80}
K{\=o}saku Yosida.
\newblock \emph{Functional Analysis}.
\newblock Classics in Mathematics. Springer, Berlin, 1995.
\newblock URL \url{https://doi.org/10.1007/978-3-642-61859-8}.
\newblock Reprint of the 6th (1980) edition [Springer, Berlin - New York].

\bibitem[Zienkiewicz et~al.(2013)Zienkiewicz, Taylor, and Zhu]{ztz:fem:13}
O.~C. Zienkiewicz, R.~L. Taylor, and J.~Z. Zhu.
\newblock \emph{The Finite Element Method: its Basis and Fundamentals}.
\newblock Elsevier/Butterworth Heinemann, Amsterdam, 7th edition, 2013.
\newblock URL \url{https://doi.org/10.1016/B978-1-85617-633-0.00001-0}.

\end{thebibliography}

\end{document}